\documentclass[letterpaper,aps,showpacs,floatfix,10pt,prc]{revtex4-1}
\usepackage{graphicx}
\usepackage{amsmath}
\usepackage{graphicx}
\usepackage{comment}
\usepackage{float}

\begin{document}
\title{Relating Measurable Correlations in Heavy Ion Collisions to Bulk Properties of Equilibrated QCD Matter}
\author{Scott Pratt}
\affiliation{Department of Physics and Astronomy and National Superconducting Cyclotron Laboratory\\
Michigan State University, East Lansing, MI 48824~~USA}
\author{Clint Young}
\affiliation{Department of Physics and Astronomy and National Superconducting Cyclotron Laboratory\\
Michigan State University, East Lansing, MI 48824~~USA}
\date{\today}

\pacs{}

\begin{abstract}
In order to compare theoretical calculations of thermal fluctuations of conserved quantities, such as charge susceptibilities or the specific heat, to experimentally measured correlations and fluctuations in heavy ion collisions, one must confront the reality of changing conditions within the collision environment, and transport of conserved quantities within the finite duration of the expansion and dissolution of the reaction. In previous work, fluctuations of conserved charges from lattice calculations, where charge is allowed to fluctuate within the designated volume consistent with the grand canonical ensemble, was linked to correlations in heavy-ion collisions, which accounted for the finite time with which to transport absolutely conserved quanatities. In this case details of the correlations were related to the evolution of the susceptibility. In this work, this paradigm is extended to compare fluctuations of momentum or energy to transverse energy correlations that can be measured in heavy-ion collisions. The sensitivity of these correlations to the equation of state, viscosity and diffusion is illustrated by considering simple models without transverse expansion. Only correlations in relative spatial rapidity are discussed here, but the prospects for extending these ideas to realistic calculations and for making realistic connections with experiment are discussed.
\end{abstract}

\maketitle

\section{Introduction}
\label{sec:intro}
Fluctuations are essential for characterizing and describing states of matter. For example, energy fluctuations are related to the specific heat which describes the equation of state, and charge fluctuations test the validity of microscopic quasi-particle descriptions.  Fluctuations can involve any conserved quantity, and are of the form $\langle \delta Q_a\delta Q_b\rangle$, and because defining conserved charges is independent of basis, are well-posed observables. Here, ``charge'' could refer to any conserved quantity, including energy or momentum. In grand canonical descriptions, where the system is assumed to be connected to both a heat bath and to a particle/charge bath, the charges within a fixed volume fluctuate. Further, if the system is comprised of independent quasi-particles, the fluctuations become Poissonian in nature and only exist between the charges on a given quasi-particle. For highly correlated systems, such as what one encounters near a phase transition, fluctuations become large and non-trivial, especially near a critical point, and for a first-order phase transition can even encompass the entire macroscopic volume if the system is in the phase coexistence region.

High-energy heavy ion collisions create environments with both novel degrees of freedom and perhaps phase discontinuities as one transitions into the Quark Gluon Plasma (QGP). Thus, fluctuations represent a crucial observable for understanding the QGP. Numerous fluctuations of quantities have been calculated for such systems, most notably with lattice gauge theory. However, such grand canonical calculations represent an idealized limit where conserved quantities are free to enter and leave the defined volume and equilibrate over infinite time. In contrast, charges within the net experimental volume are fixed and do not fluctuate. Within a sufficiently small sub-volume, charges have sufficient time to enter and leave to approach a grand-canonical limit, but defining that volume is problematic. As the system expands and cools, so do the equilibrated fluctuation observables. Thus, coming to equilibrium represents hitting a moving target, and even defining a scale for such a sub-volume is questionable at best. 

Fortunately, correlations binned by relative rapidity, azimuthal angle or momentum, can provide the link between fluctuations and susceptibilities in the grand canonical ensemble to correlations that are meaningful and measurable in the context of heavy-ion collisions. Such correlations are expressed as $\langle\delta\rho_a(\vec{r}_1,t)\delta\rho_b(\vec{r}_2,t)\rangle$ where $\rho_a$ refers to a density of a conserved charge and $x_i$ refer to some measure of position. In this work correlations are considered only to the level of two-point functions, but in principle one could consider $n>2$ point correlations. In \cite{Pratt:2012dz,Pratt:2015jsa} the connection between correlations and fluctuations was demonstrated for the fluctuation of up, down and strange charges (or equivalently electric charge, baryon number and strangeness). It was also shown how these correlations, labeled up-down-strange, project onto correlations binned by relative rapidity for specific hadron species, referred to as generalized charge balance functions. Within a rather simple model that parameterized the diffusion of charges, the data suggested that matter in heavy-ion collisions comes within a few tens of percent of chemical equilibration as defined by the charge susceptibilities from lattice calculations, within the first $\sim 1$ fm/$c$ of the collision.

The purpose of this work is to establish a paradigm for connecting any two-point correlation, susceptibility or fluctuation that can be formed from conserved quantities to measurable correlations in the final state. This general framework forms the phenomenological infrastructure for meaningful comparisons between lattice gauge theory and experiment. In addition to fluctuations and susceptibilities, measured correlations are also shown to be sensitive to transport coefficients such as the diffusion constant or the viscosity. These transport coefficients describe how energy, momentum or charges dissipate throughout the system. In the next section the methodology is presented in general terms for connecting the fluctuation of any types of charge to the corresponding correlation function, using the method of \cite{Pratt:2012dz} for up/down/strange charge as an example. In some ways these methods are similar to those in \cite{Gavin:2006xd,Gavin:2012if}, but the treatment across different types of correlations is emphasized. Sections \ref{sec:px} and \ref{sec:et} present simple calculations for correlations of transverse momentum, $p_x$ or $p_y$, and transverse energy respectively. Section \ref{sec:et} is especially lengthy due to the fact that energy can move hydrodynamically and its evolution is intertwined with momentum correlations. Section \ref{sec:cascade} describes how to perform the last step, projecting correlation functions of specific charges into correlations indexed by specific hadron species. Results are summarized in \ref{sec:summary} and the prospect for using measured correlations to infer information about fluctuations and transport coefficients in an equilibrated system is evaluated.

\section{General Method}
\label{sec:general}

In a large volume that has equilibrated over an infinitely long time, and in the absence of a phase transition, correlations have a finite range and can be expressed as
\begin{equation}
C_{ab}(\vec{r}_a,\vec{r}_b)
=\langle\delta\rho_a(\vec{r}_a)\delta\rho_b(\vec{r}_b)\rangle=\int d\vec{r}~\chi_{ab}\delta(\vec{r}_a-\vec{r}_b).
\end{equation}
where $\vec{r}_a$ refers to the position in coordinate space. The coordinate could refer to spatial rapidity, or could be multidimensional, e.g. referring to $\vec{r}_a$. The fluctuation $\delta\rho_a=\rho_a-\langle\rho_a\rangle$ refers to a conserved quantity. Here $\delta(\vec{r}_a-\vec{r}_b)$ is a short-range function that integrates to unity. For most purposes one can consider it to be a true delta function, but it doesn't change the subsequent derivations if it has a finite  but small extent. 

Using the definition of $\chi$ above, the charge fluctuation, $\delta Q_a=\int d\vec{r}_a\delta\rho_a$, becomes
\begin{eqnarray}
\frac{1}{\Omega}\langle \delta Q_a\delta Q_b\rangle&=& \frac{1}{\Omega}\int_\Omega d\vec{r}_ad\vec{r}_b~ C_{ab}(\vec{r}_a-\vec{r}_b)=\chi_{ab}.
\end{eqnarray}
Here $\Omega$ is the generalized volume, $\int d\vec{r}=\Omega$. The expression above expresses little more than the assumption that correlations that lead to non-zero fluctuations for an equilibrated system are all local.

For the gaseous limit, there exist well-defined quasi-particles, and particles become uncorrelated with one another. In that limit correlations are only those within the quasi-particle. The susceptibility then becomes
\begin{equation}
\chi_{ab}=\sum_i n_i q_{ia}q_{ib},
\end{equation}
where $q_{ia}$ is the charge of type $a$ for species $i$ with density $n_i$. For example, if one had a gas of protons, the susceptibility for $uE$, referring to the {\it up} charge and energy, would be
\begin{equation}
\chi_{uE}=\frac{(2S+1)}{(2\pi)^3}\int d^3p~e^{-(E_p-\mu)/T} (2\cdot E_p),
\end{equation}
where the factor $(2\cdot E_p)$ comes from the product of the up charge (2) and the energy $E_p$. The spin of the proton here is $S$, and we will ignore corrections from identical particle statistics, which should be negligible for all particles besides pions in the environment of a heavy-ion collision.

If the volume is finite, and if total charge is conserved, the situation is quite different. Each integral of either the charge density or correlation goes to zero.
\begin{eqnarray}
\label{eq:chargecons}
\int d\vec{r}_a\delta\rho_a&=&0,\\
\label{eq:chargecons2}
\int d\vec{r}_b~C_{ab}(\vec{r}_a,\vec{r}_b)&=&0.
\end{eqnarray}
Assuming the system has time to locally equilibrate, the correlations at small relative coordinate can be expected to approach the equilibrated value. The correlation can then be decomposed into two parts,
\begin{eqnarray}
\label{eq:Cprimedef}
C_{ab}(\vec{r}_a,\vec{r}_b)&=&\chi_{ab}(\vec{r}_a)\delta(\vec{r}_a-\vec{r}_b)
+C'_{ab}(\vec{r}_a,\vec{r}_b),\\
\nonumber
\int d\vec{r}_b ~C'_{ab}(\vec{r}_a,\vec{r}_b)&=&-\chi_{ab}(\vec{r}_a). 
\end{eqnarray}

The function $C'$ describes the balancing charge, or energy or momentum, which for a large static system would spread out over an increasingly large sub-volume as time progresses. If allowed infinite time to equilibrate in a large system, the magnitude of $C'$ would disappear even though the integrated value would remain at $-\chi$. 

For an expanding and cooling system, $\chi$ evolves with time, and so does $C'$ For each type of charge $a$ used to construct $\chi$, there is an equation to express local conservation of charge,
\begin{equation}
\label{eq:cont}
\partial_t\delta\rho_a(r,t) +\nabla\cdot\delta\vec{j}_{a}(r,t)=0, 
\end{equation}
where $\delta \vec{j}_a$ is defined by the dynamics, and can be thought of a as generalized current. For example, if $\rho_a$ in Eq. (\ref{eq:cont}) refers to a charge density and if the charge evolves diffusively, $\delta \vec{j}_a(r)=-D\nabla\delta\rho_a(r)$ where $D$ is the diffusion constant. If one were in Bjorken coordinates \cite{bjorken:83}, the gradient would be $(1/\tau)\partial_{\eta,a}$ with $\tau$ being the proper time, $\tau=\sqrt{t^2-z^2}$, and $\eta$ being the spatial rapidity, $\eta=\tanh^{-1}z/t$. The derivatives $\partial_t$ would be replaced by $\partial_\tau$. If $\rho_a$ is an energy density $\vec{j}_a$ is the momentum density appropriate for the dimensionality.

The evolution of $C'_{ab}$ then becomes
\begin{eqnarray}
\label{eq:Sdef}
\partial_tC'_{ab}(\vec{r}_a,\vec{r}_b,t)&=&-\nabla_a\cdot\langle \delta \vec{j}_a(\vec{r}_a,t)\delta \rho_b(\vec{r}_b,t)\rangle
-\nabla_b\cdot\langle\delta\rho_a(\vec{r}_a,t)\delta\vec{j}_b(\vec{r}_b,t)\rangle\\
\nonumber
&&+S_{ab}(\vec{r}_a)\delta(\vec{r}_a-\vec{r}_b).
\end{eqnarray}
The source function $S_{ab}$ accounts for the fact that $C'$ neglects that part of the correlation from $\vec{r}_a\approx\vec{r}_b$. A form for $S_{ab}$ is determined by charge conservation, Eq. (\ref{eq:chargecons2}) and the definition of $C'$ in Eq. (\ref{eq:Cprimedef}). 
\begin{eqnarray}
\label{eq:Cprimeevolve}
&&\hspace*{-60pt}\int d\vec{r}_ad\vec{r}_b~\left\{\partial_tC'_{ab}(\vec{r}_a,\vec{r}_b,t)+\nabla_a\cdot\langle \delta \vec{j}_a(\vec{r}_a,t)\delta \rho_b(\vec{r}_b,t)\rangle+\nabla_{b}\langle\delta\rho_a(\vec{r}_a,t)\delta\vec{j}_b(\vec{r}_b,t)\rangle\right\}\\
\nonumber
&=&-\partial_t\int d\vec{r}_ad\vec{r}_b~\chi_{ab}(\vec{r_a},t)\delta(\vec{r}_a-\vec{r}_b)\\
\nonumber
&=&-\partial_t \int d\vec{r}~\chi_{ab}(\vec{r})=\int d\vec{r}~S_{ab}(\vec{r},t).
\end{eqnarray}
Susceptibilities are defined as the fluctuation in a fixed volume. In an expanding or flowing system the volume of a fluid element can expand or contract over time. For a volume element $d\Omega$
\begin{eqnarray}
\nonumber
d\Omega~ S_{ab}(r,t)&=&-\partial_t\left(d\Omega ~\chi_{ab}(r,t)\right)\\
\nonumber
&=&-d\Omega~\left(\partial_t\chi_{ab}+\chi_{ab}\nabla\cdot\vec{v}\right)\\
\nonumber
&=&-d\Omega~\partial_\mu(u^\mu\chi_{ab}),\\
\label{eq:truesource}
S_{ab}(r,t)&=&-\partial_\mu(u^\mu\chi_{ab}).
\end{eqnarray}
Here, $u$ is the four-velocity in the fluid frame, so $\partial_t=u\cdot\partial$ and $\nabla\cdot v=\partial\cdot u$. The source function $S_{ab}$ changes as the susceptibility changes. The contribution from $\nabla\cdot v$ can be replaced if one assumes entropy conservation,
\begin{eqnarray}
\nonumber
\partial_\mu(u^\mu s)&=&0,\\
\nonumber
\partial\cdot u&=&-\frac{(u\cdot\partial) s}{s},\\
S_{ab}(r,t)&=&-s(u\cdot \partial)\left(\frac{\chi_{ab}}{s}\right).
\end{eqnarray}
This emphasizes that the sign of the source function is largely driven by whether $\chi/s$ rises or falls as the system expands. If entropy is not conserved to the desired accuracy, one merely reverts to Eq. (\ref{eq:truesource}).

For the Bjorken case with no transverse expansion, where averaged quantities depend only on the proper time $\tau$ and the coordinate $r$ refers to spatial rapidity, it is convenient to use quantities that refer to densities per spatial rapidity, rather than densities per Cartesian volume. The susceptibility $\chi_{ab}$ refers to a fluctuation per unit rapidity, and if one assumes that the entropy per unity spatial rapidity is constant, the fluctuation per rapidity, $\chi_{ab}$ is related to the fluctuation per Cartesian volume, which is the lattice value, by
\begin{equation}
\chi_{ab}(\eta,\tau)=\frac{dS}{d\eta}\left(\frac{\chi_{ab}^{\rm(latt)}(\tau)}{s^{\rm(latt)}(\tau)}\right),
\end{equation}
where $dS/d\eta$ is the rapidity per spatial rapidity which is approximately constant. The source function when the coordinates are the spatial rapidity is then
\begin{eqnarray}
S_{ab}(\eta,\tau)&=&\frac{dS}{d\eta}\partial_\tau\left(\frac{\chi^{\rm(latt)}_{ab}(\eta)}{s^{\rm(latt)}}\right).
\end{eqnarray}
Thus, even though one may be working is a Bjorken coordinates, the ratio of $\chi/s$ can still be constructed with $\chi$ and $s$ referring to number per Cartesian volume, and thus taken directly from lattice. 

For the fully three-dimensional case with transverse expansion included, one can follow similar steps. One would begin writing
\begin{eqnarray}
\partial_\tau \delta\rho_a(\tau,x,y,\eta)+\partial_i j_{a,i}(x,y,\eta,\tau)&=&0,\\
\partial_i&=&(\partial_x,\partial_y,(1/\tau)\partial_\eta).
\nonumber
\end{eqnarray}
Following the steps above and still assuming conservation of entropy, one finds
\begin{eqnarray}
\label{eq:sourcedchidt}
\partial_\tau C'_{\tau,ab}(\vec{r}_a,\cdots\vec{r}_b)&+&\partial_i\langle \delta j_{a,i}(\tau,\vec{r}_a)\delta \rho_b(\tau,\vec{r}_b)\cdots \rho_f(\tau,\vec{r}_b)\rangle+\cdots
+\partial_i\langle\delta\rho_a(\tau,\vec{r}_a)\delta j_{b,i}(\tau,\vec{r}_b)\rangle\\
\nonumber
&=&S_{ab}(\vec{r}_a)\prod_{i=b\cdots c}\delta(x_i-x_a)\delta(y_i-y_a)\frac{1}{\tau}\delta(\eta_i-\vec{r}_a),\\
\nonumber
S_{ab}(\tau,x,y,\eta)&=&-s(\tau,x,y,\eta)u\cdot\partial
\left(\frac{\chi^{\rm(latt)}_{ab}(x,y,\eta,\tau)}{s^{\rm(latt)}(x,y,\eta,\tau)}\right).
\end{eqnarray}

\subsection{Implementation Strategies}

Once one has chosen the correlation of interest and identified the appropriate source function, one must solve the equations. This can be done in one of two strategies. First, one could solve Eq. (\ref{eq:Cprimeevolve}) for the evolution of $C'_{ab}(\vec{r}_a,\vec{r}_b,t)$. This may be difficult because the evolution involves both $\vec{r}_a$ and $\vec{r}_b$. It also considers only the case at equal times, $t_a=t_b=t$. For some cases the more reasonable strategy is to evolve the charges of type $a$ and $b$ separately. Consider a point charge
\begin{equation}
\delta\rho_a(\vec{r},t_a)=Q_a\delta(\vec{r}-\vec{r}_a),
\end{equation}
whose evolution is defined by a propagator $G$,
\begin{equation}
\delta\rho_c(r_c)=G_{ca}(r_c,r_a)Q_a,
\end{equation}
where $r$ refers to the four-vector.The off-diagonal elements occur when a charge of type $a$ inspires a change in a charge of type $c$. The four-vectors $r_c$ and $r_a$ incorporate the time components. For instance, in hydrodynamics a perturbation of the energy density affects the momentum density. For non-zero baryon density, a perturbation of energy density affects the baryon density and vice-versa.

Once one solves for the propagator, one can find the correlations, even at non-equal times, given the source function,
\begin{eqnarray}
\langle \delta\rho_c(r_c)\delta\rho_d(r_d)\rangle'&=&
\int d^4r~S_{ab}(r)G_{ca}(r_c-r)G_{db}(r_d-r).
\end{eqnarray}
Again, the prime denotes that the correlation neglects the contribution from $r_c\approx r_d$.

In practice performing such an integral would probably be done with Monte Carlo methods, by randomly generating discrete sources at points $\vec{r}$, such that the probability of generating a point within $d^4r$ was $\kappa rS_{ab}(r)d^4r$. One would consider the impact of a unit charge $a$ beginning at $r$. One would then do the same for charge $b$, then consider the binning of the correlation $C_{cd}(r_a,r_d)$ according to the charges that appear in final bins $r_c$ and $r_d$. The binning would then be weighted by $1/\kappa$,  with $\kappa$ chosen to provide a sufficiently enough number of points for good sampling, but not so large as to be computationally prohibitive. This method is built on an implicit assumption, that the charges evolve independently, with their correlation deriving solely from the source function $S_{ab}(r)$. This is valid in the limit that the perturbations $\delta\rho$ are small and evolve linearly. 

\subsection{Susceptibilities and their relation to partition functions}

\begin{table}
\begin{tabular}{|r|c|l|}
\hline
$\langle \delta Q_a\delta Q_b\rangle$ &\parbox{3.6in}{$3\times 3$ matrix, off-diagonal terms in hadronic phase} & $=T^2\partial^2/\partial \mu_a\partial\mu_b\ln Z$\\
\hline
$\langle \delta Q_a\delta E\rangle$ & \parbox{3.6in}{would vanish unless average charge $\langle Q_a\rangle\ne 0$}&
$=-\partial^2/\partial\mu_a\partial\beta\ln Z +(\partial/\partial\mu_a\ln Z)(\partial/\partial\beta\ln Z)$\\
\hline
$\langle \delta E\delta E\rangle$& \parbox{3.6in}{energy fluctuations are related to specific heat} 
& $=\partial^2/\partial\beta^2\ln Z=-\partial E/\partial\beta=T^2C_V$\\
\hline
$\langle \delta p_i\delta p_i\rangle$& \parbox{3.6in}{momentum fluctuations are determined by inertial mass} 
& $=(P+\epsilon)VT =TV\ln Z-V\partial/\partial\beta\ln Z$\\
\hline
\end{tabular}
\caption{\label{table:chiab}Examples of susceptibilities $\chi_{ab}$}
\end{table}

Correlations of conserved quantities can be generated by taking two partial derivatives of the grand canonical partition function $Z(\mu,T)$, where the temperature $T$ and the chemical potentials $\mu_a$, one for each conserved charge, are sufficient to define the partition function. A list of potential quantities from which to construct correlators binned by rapidity are listed in Table \ref{table:chiab}. Energy fluctuations are related to the specific heat at a finite volume $V$, 
\begin{eqnarray}
C_V&=&\frac{\partial}{\partial T}\langle E\rangle/V=T^2 \partial_\beta^2(\ln Z/V).\\
\nonumber
&=&\frac{1}{T^2}\left(\langle E^2\rangle-\langle E\rangle^2\right)/V=\frac{1}{T^2}\langle \delta E^2\rangle/V.
\end{eqnarray}
Charge fluctuations come from taking derivatives of $\ln Z$ with respect to the chemical potentials,
\begin{eqnarray}
\chi_{ab}&=&\langle \delta Q_a\delta Q_b\rangle/V\\
\nonumber
&=&T^2\frac{\partial}{\partial\mu_a}\frac{\partial}{\partial\mu_b}(\ln Z/V).
\end{eqnarray}
Whereas the expressions above are commonly found in textbooks, it is less likely to find the expression for the fluctuation of the momentum. For instance, the $x$ component of the momentum non-relativistically would have fluctuations determined by the equipartition theorem. 
\begin{eqnarray}
\label{eq:inertialmass}
\langle\delta p_x\delta p_x\rangle&=&MT\\
\nonumber
&=&(P+\epsilon) VT.
\end{eqnarray}
Here, the mass of the system is the inertial mass density $(P+\epsilon)$ multiplied by the volume $V$. To see that $P+\epsilon$ is indeed the inertial mass density one can consider the stress energy tensor elements for small collective velocities $v$,
\begin{eqnarray}
T_{0i}&=&(P+\epsilon) v_i,\\
\nonumber
p_i&=&T_{0i}V=(P+\epsilon)Vv_i.\\
\nonumber
T_{00}&=&\epsilon+(P+\epsilon)v^2/2\\
\nonumber
E&=&T_{00}V=\epsilon V+\sum_i\frac{p_i^2}{2(P+\epsilon)V}.
\end{eqnarray}
Thus, the equipartion theorem leads to Eq. (\ref{eq:inertialmass}).

In the vicinity of a phase change, susceptibilities tend to peak, and can become singular in the limit of a phase transition. As the system proceeds through the region where the susceptibility rises and falls, source functions  correspondingly turn negative and positive, i.e. tend negative as the system enters the transition region, followed by turning positive at later times when the fluctuation is diminishing relative to the entropy. The switching of the sign of the source function can result in non-trivial shapes to the correlations in the final state.

The charges that might be referenced are up, down and strange, or equivalently electric charge, baryon number and strangeness. Conservation of energy and momentum provide $E$, $P_x$, $P_y$ and $P_z$ as indices. Any two of the seven possibilities can be combined for a susceptibility. However, due to symmetry the fluctuation of $P_i$ with any of the other six quantities would be zero. At zero average baryon number $\langle \delta B\delta E\rangle=0$, but at lower energy where the mean baryon number is significant, this quantity can also play a role. Asymptotically, any particle's final spatial rapidity $\eta$ will approach its regular rapidity, $y=(1/2)\ln[(1+v_z)/(1-v_z)]$, and because rapidity is used as a proxy for spatial position, it cannot fluctuate relative to its position. Therefore, any correlator with $\delta p_z$, where $p_z$ is measured relative to a frame moving with the average velocity of the local matter, will vanish. One may need correlators involving $\delta p_z$ for calculating the evolution of other correlators, but they should vanish at large time. Asymptotically at a given $\eta$, energies are all transverse energies because any particle's trajectory approaches the bin of spatial rapidity $\eta$ defined by its velocity, and the measurement of a particle's longitudinal momentum relative to a frame moving with a velocity $v_z=z/t$ approaches zero at large time. 

\begin{figure}
\centerline{\includegraphics[width=0.47\textwidth]{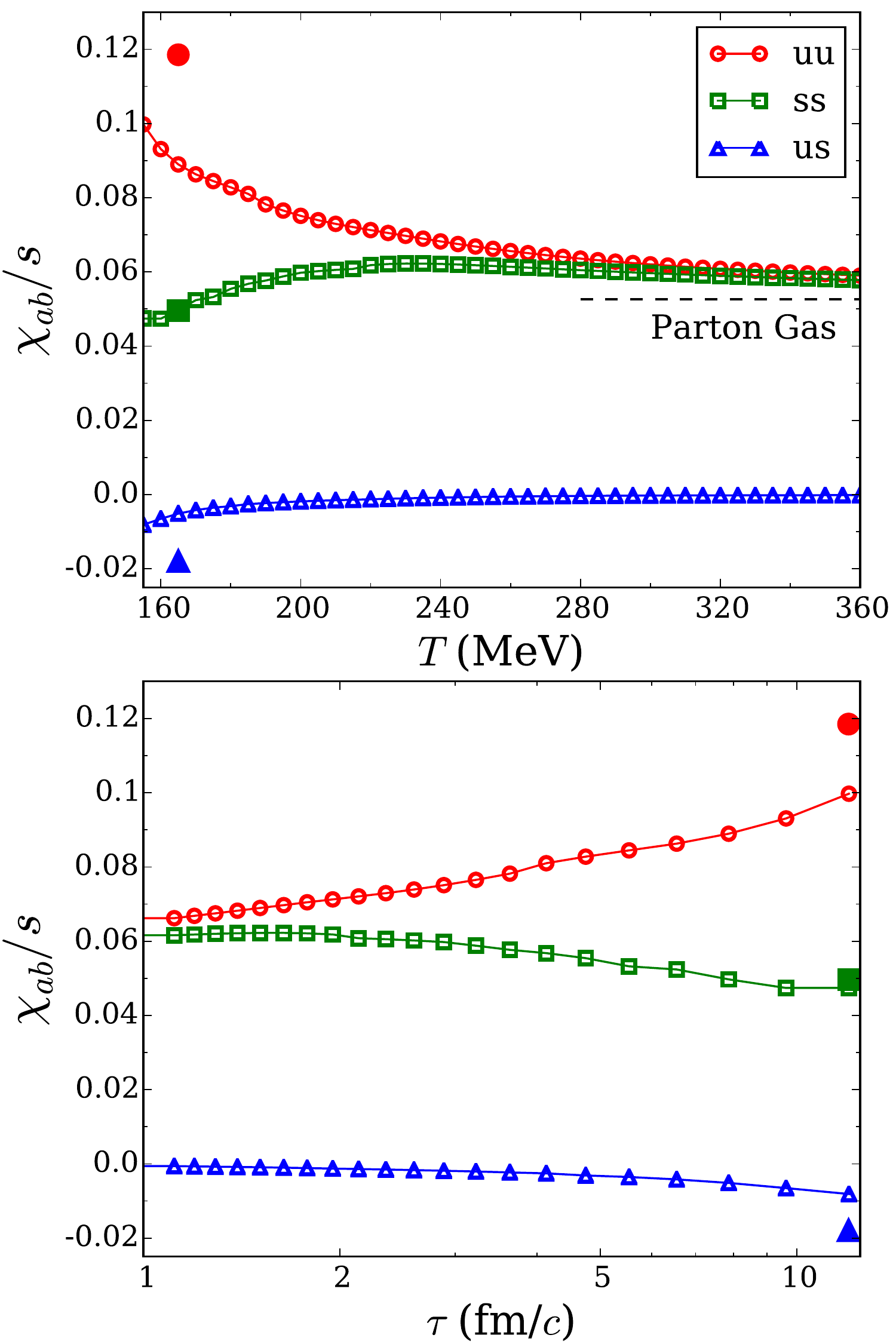}}
\caption{\label{fig:latticechargefluc}(color online) 
The source for generating correlations, see Eq. (\ref{eq:sourcedchidt}), is driven by $S\sim -\partial_\tau\chi_{ab}/s$. For charge correlations $\chi/s$ is shown in the upper panel from lattice gauge theory \cite{Borsanyi:2011sw,Bellwied:2015lba} for the up-up, strange-strange and up-strange charge fluctuations. The three filled points show susceptibilities for a hadron gas at $T=165$ MeV. Assuming a purely one-dimensional Bjorken expansion, temperature vs. time graphs were generated using the lattice equation of state. The lower panel shows the same susceptibilities plotted against time. The source function for up-up correlations has an initial surge corresponding to $\chi/s$ reaching its equilibrated alue at $\tau\approx 1$ fm/$c$. After than time there is a steady rise of $\chi_{uu}/s$ corresponding to a steady source function with a strengthening as one approaches $T_c$. In contrast the strange-strange susceptibility slightly weakens as one cools and the later-stage source function has the opposite sign.
}\end{figure}
Charge fluctuations are the easiest to model, because the charge moves diffusively in the limit of zero baryon density, which allows transport to be treated as a random walk. Figure \ref{fig:latticechargefluc} shows how $\chi/s$ behaves with temperature and time. To crudely project $\chi/s$ vs $T$ into a $\chi/s$ vs $\tau$ plot, the temperature vs. time plot was generated assuming an ideal one-dimensional Bjorken expansion, with the temperature of 155 MeV being reached at 12 fm/$c$, and the initial temperature at $\tau=1$ fm/$c$ being 265 MeV. There is a rise in $\chi_{uu}/s$ and $\chi_{dd}/s$ as one cools into the transition region. In contrast $\chi_{ss}/s$ modestly drops as one enters the transision region. This corresponds to a sudden negative contribution to the source function, which manifests itself in non-trivial behavior of charge-balance correlations. References \cite{Pratt:2015jsa} show how these features can be exploited to constrain the chemical make-up of the early stages of the collision. Transverse momentum fluctuations are related to experiment by measuring the momentum-momentum correlations of given bins in final-state relative rapidity. These measurements can be performed for both in-plane and out-of-plane momenta. In \cite{Pratt:2010zn} this difference was shown to be important in measurements of angular correlations involved in searching for the chiral magnetic effect \cite{starcme}. In the next subsection, it will be shown how the $\langle\delta p_x\delta p_x\rangle$ correlation binned by relative rapidity widens with time according to a diffusion equation with the shear viscosity determining the diffusion constant. Transverse energy correlations are the subject of the next section. Due to the way that the equations of motion for the stress-energy tensor mix momentum and energy, the evolution of these correlations, and even the appropriate definition of such correlations is tricky. Compared to the transverse momentum correlations and the charge-charge correlations, the formalism is much more involved and requires understanding of subtle details.


\section{Transverse Momentum Fluctuations}
\label{sec:px}

As listed in Table \ref{table:chiab} the susceptibility for transverse momentum fluctuations is a product of the enthalpy, $P+\epsilon$, and the temperature,
\begin{eqnarray}
C_{xx}(\eta)&=&\langle \delta P_x(0) \delta P_x(\eta)\rangle,\\
\nonumber
&=&(P+\epsilon)T\pi R^2\tau\delta(\eta)+C'_{xx}(\eta),\\
P_x(\eta)&\equiv&\tau\int dxdy ~T_{0x}(x,y,\eta,\tau).
\end{eqnarray}
Here, $P_x$ is the net momentum in the $x$ direction in a small bin of spatial rapidity $\eta$ divided by the bin width. The equations of motion for $\delta P_x$ are generated through energy-momentum conservation applied to the stress-energy tensor, $\partial_\mu T^{\mu\nu}=0$.
\begin{eqnarray}
\partial_\tau \delta P_x&=&-\tau\int dxdy~\left(\partial_x\delta T_{xx}+\partial_y\delta T_{xy}+(1/\tau)\partial_\eta \delta T_{xz}-(1/\tau)\delta T_{0x}\right)\\
\nonumber
&=&-\partial_\eta\int dxdy~\left(\delta T_{xz}(x,y,\eta,\tau)-\eta\delta T_{0x}(x,y,\eta,\tau)\right)\\
\nonumber
&=&
-\partial_\eta\int dxdy~\delta \tilde{T}_{xz}(x,y,\eta,\tau).
\end{eqnarray}
Here, $\delta \tilde{T}$ refers to the the stress-energy tensor as defined in the frame moving with rapidity $\eta$. For small $\eta$, $T_{xz}=\tilde{T}_{xz}+\eta T_{0x}$. One can then use the Navier Stokes equation to state
\begin{eqnarray}
\partial_\tau \delta P_x&=&\eta_s\partial_\eta\int dxdy~ \frac{1}{\tau}\partial_\eta \delta v_x(x,y,\eta,\tau)\\
\nonumber
&=&\frac{\eta_s}{\tau}\partial_\eta\int dxdy~\partial_\eta \left(\frac{\delta T_{0x}(x,y,\eta,\tau)}{P+\epsilon}\right).
\end{eqnarray}
Here, we have assumed viscous corrections are small with $\eta_s$ denoting the shear viscosity, and that the evolution is described by Navier Stokes equations. Transverse flow is neglected and assuming that matter uniform in the transverse direction, 
\begin{eqnarray}
\label{eq:Dxx}
\partial_\tau \delta P_x&=&\frac{D}{\tau^2}\partial_\eta^2\delta P_x,\\
\nonumber
D&=&\frac{\eta_s}{P+\epsilon}.
\end{eqnarray}
Thus, transverse momentum fluctuations spread diffusively, with the diffusion constant being determined by the ratio of the shear viscosity to the enthalpy, a result shown in \cite{Pratt:2010zn}. One can then consider the correlation $C'_{xx}(\eta_2-\eta_1,\tau)$. The equations of motion for $C'_{xx}$ are
\begin{eqnarray}
\label{eq:Sxx}
\partial_\tau C'_{xx}(\eta,\tau)&=&\frac{2D}{\tau^2}\partial_\eta^2C'_{xx}(\eta,\tau)+S_{xx}(\tau)\delta(\eta),\\
\nonumber
S_{xx}(\tau)&=&-\pi R^2\frac{d}{d\tau}\left[(P+\epsilon)T\tau\right]\\
\nonumber
&=&-\frac{dS}{d\eta}\frac{d}{d\tau} T^2.
\end{eqnarray}
The last step assumed that the conserved entropy per unit rapidity was $dS/d\eta=(P+\epsilon)\pi R^2\tau/T$, which would be true for the zero net charge limit.

Because the temperature falls with time, the source term $S_{xx}$ stays positive for any equation of state. However, varying the equation of state clearly affects the source term, and therefore changes $C'_{xx}$. For example, if one believed in limiting temperatures, i.e. the Hagedorn model \cite{hagedorn}, the source term would tend to zero as one approached $T_{\rm max}$ and $C'_{xx}$ would mainly have contributions from the earliest times from when the matter was thermalizing. These correlations which would then diffuse and become relatively weak by breakup. The strength of this initial correlation would correspond to $T^2$ rising from zero to the equilibrated value during equilibration. Once thermalized, there would be no additional source contributions until the temperature fell below $T_{\rm max}$.

\begin{figure}
\centerline{\includegraphics[width=0.45\textwidth]{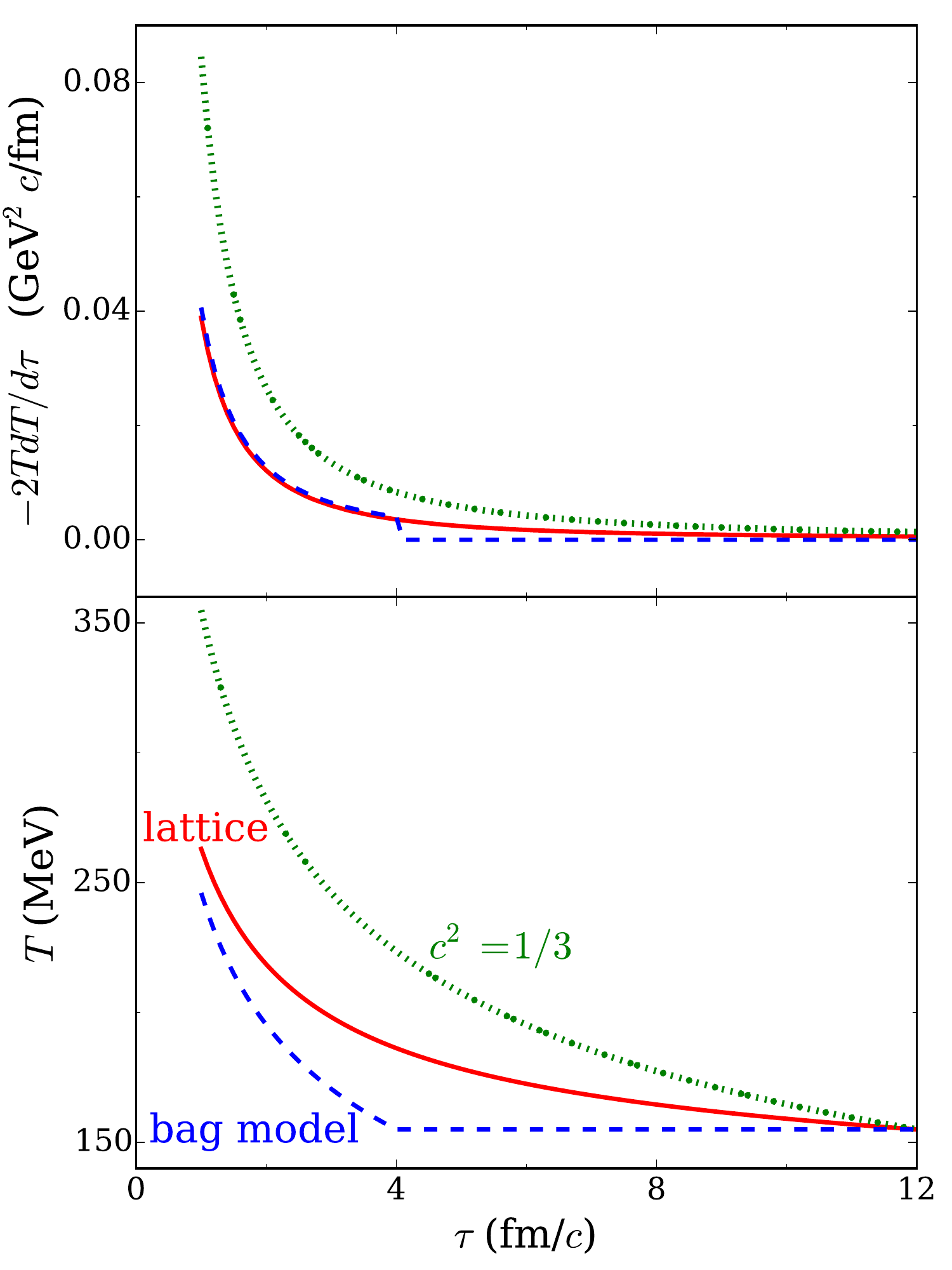}}
\caption{\label{fig:Tvstau}(color online) 
The temperature vs time is shown for three equations of state in the lower panel where the matter undergoes a one-dimensional Bjorken expansion and reaches a temperature $T=155$ MeV at $\tau=12$ fm/$c$. The lattice equation of state is softer than that of a gas of massless partons, where the speed of sound is a constant $c_s^2=1/3$, and is stiffer than a bag-model equation of state, which features a strong first-order phase transition. The upper panel shows $TdT/d\tau$, which drives the source function for transverse momentum correlations, as seen in Eq. (\ref{eq:Sxx}). The stiffer equations of state have significantly stronger sources throughout the evolution, which should lead to stronger correlations of transverse momentum at lower relative rapidity. }\end{figure}

Stiff equations of state are characterized by relatively rapidly falling temperatures, which then provide positive contributions to $C'_{xx}$ at all times which would have little time to diffuse and weaken. Figure \ref{fig:Tvstau} shows both the temperature and $-TdT/d\tau$, which describes the source function in Eq. (\ref{eq:Sxx}), as a function of time for an isentropic one-dimensional boost-invariant expansion. Functions are shown for three equations of state, with each having reached $T=155$ MeV at $\tau=12$ fm/$c$. Temperatures at earlier times depend on the equation of state and are determined by assuming an ideal expansion with the entropy density scaling as $1/\tau$. The first equation of state considered is the lattice equation of state, and is compared to a much stiffer equation of state with constant speed of sound, $c^2=1/3$, and is also compared to a bag model equation of state, which is softer. In the bag model, a mixed phase is assumed until the matter is heated to the point where the hadron gas is completely transformed into a massless gas of gluons and up, down and strange quarks. For the softer equations of state the temperatures are lower at a given time and the source functions, which scale as $-TdT/d\tau$, are significantly reduced, especially at later times.

Correlations were calculated for this simplified case with boost invariance assumed and transverse flow neglected. The diffusion equation was treated numerically by a random walk. Sample particles transported transverse momentum with random velocities uniformly chosen between $\pm$ the speed of light relative to the local fluid. Random walks provide solutions to the diffusion equation in the limit of many steps if the variance of the step size in the longitudinal direction satisfies the relation
\begin{equation}
\langle \delta z^2\rangle = 2D\delta\tau,
\end{equation}
where the mean time between random steps is $\delta\tau$. For steps that correspond to the random range of velocities mentioned above, $\langle \delta z^2\rangle=\delta\tau^2/3$. Using the expression for the diffusion constant in Eq. (\ref{eq:Dxx}), the mean random time step is
\begin{equation}
\delta\tau=6\frac{\eta_s}{P+\epsilon}.
\end{equation}
At each small time step $d\tau$, the particles were given a chance of reorienting their velocities with a probability $d\tau/\delta\tau$. For evolutions lasting many times beyond $\delta\tau$, one approaches solutions to the diffusion equation. Given that the evolution in these simulations is typically a few, perhaps up to ten, time steps, the resulting evolution does modestly differ from the more Gaussian form of a diffusion equation. Most notably, the tails of the distribution are truncated as transport is confined to being within the light cone. In practice, at each time step one creates pairs of particles with probability given by the source function, $-TdT$, where $-dT$ is the fall of the temperature during the time step. The particles are then transported according to the random walk outlined above, and the distribution of relative positions, $\Delta\eta=\eta_2-\eta_1$, provides the momentum correlation.

\begin{figure}
\centerline{\includegraphics[width=0.8\textwidth]{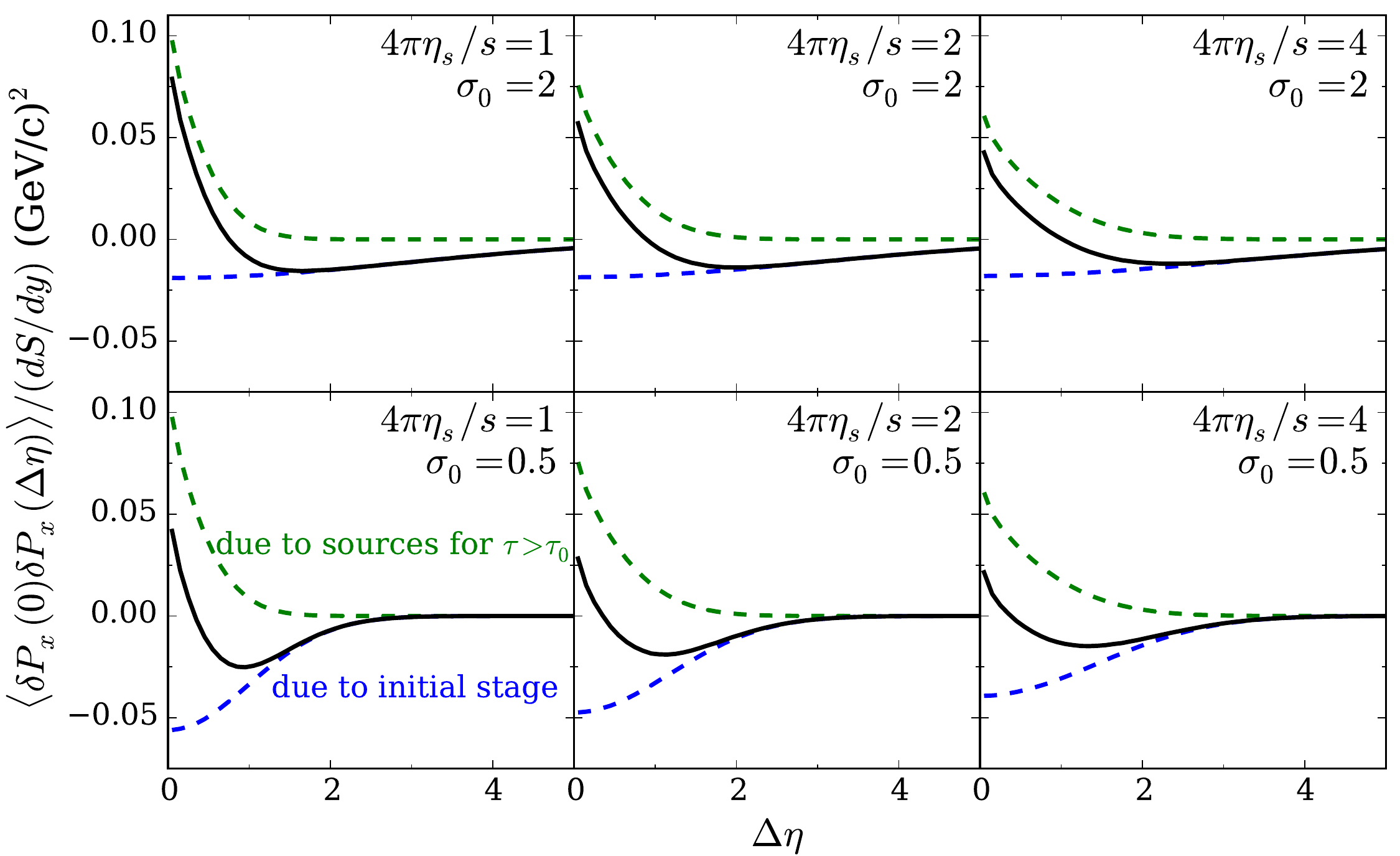}}
\caption{\label{fig:pxpxsigma0}(color online)
Transverse momentum correlations at a time $\tau_f=12$ fm/$c$ for a boost-invariant model where the temperature reached 155 MeV at $\tau_f$ and earlier temperatures were determined by entropy conservation, $s\sim 1/\tau$, combined with the lattice equation of state. The correlation spreads diffusively with the diffusion constant being determined by the shear viscosity, Eq. (\ref{eq:Dxx}), and the source coming from Eq. (\ref{eq:Sxx}). Larger viscosities broaden both the correlation from the initial state (lower blue dashed line) and that from the source function, $S_{xx}(\tau>\tau_0)$ with $\tau_0=1$ fm/$c$ (upper green dashed line). Results are also affected by the unknown width of the initial correlation at $\tau_0$. Results are shown for both narrow and broad initial widths, $\sigma_0=0.5,2.0$, and for three viscosities, $4\pi\eta_s/s=1,2$ and 4. The black lines indicate the net correlation.}
\end{figure}
The resulting transverse momentum correlation $C_{xx}(\eta=\eta_2-\eta_1)$ depends on the viscosity, which determines the rate at which the correlations spread after being created at earlier times by the source function $S_{xx}$. The correlation also must account for the initial correlation at the time at which the matter thermalizes. The initial width of this initial correlation is unknown due to our lack of knowledge of the dynamics of the pre-equilibrated matter, and it also subsequently spreads according to the same diffusive mechanism as the correlations generated by the better understood sources $S_{xx}(\tau>\tau_0)$. Figure \ref{fig:pxpxsigma0} shows correlations in a one-dimensional boost invariant system at the final time, $\tau_f=12$ fm/$c$ as generated by both the initial correlation and from $S_{xx}(\tau>\tau_0)$. Resulting correlations are shown assuming the lattice equation of state, which determines the temperature vs. time trajectory by entropy conservation, for an initial time $\tau_0=1$ fm/$c$.  The initial correlation is assumed to be Gaussian with widths of either 0.5 or 2.0 units of spatial rapidity. Viscosities of $4\pi\eta/s=1,2$ and 4 are considered in the various panels. The strength of the correlation from the initial state is fixed in strength (area above the curve) by the susceptibility, but the width and shape are not understood. This contribution spreads by approximately one unit of rapidity by $\tau=12$ fm/$c$. The initial spread of the correlation at $\tau_0$, $C_{xx}(\tau=\tau_0)$, might already be on the order of one unit of rapidity. Thus, the diffusive spreading for $\tau>\tau_0$ is not so large that one needn't worry about the initial spread. Higher viscosity leads to more spread out correlations. Quadrupling the viscosity approximately doubles the width, as expected given that the diffusion constant scales proportional to the shear viscosity. If the width of the negative correlation at large $\Delta\eta$ can be used to constrain $\sigma_0$, the shape of the narrower peak at small $\Delta\eta$ can then be used to constrain the shear viscosity.

\begin{figure}
\centerline{\includegraphics[width=0.7\textwidth]{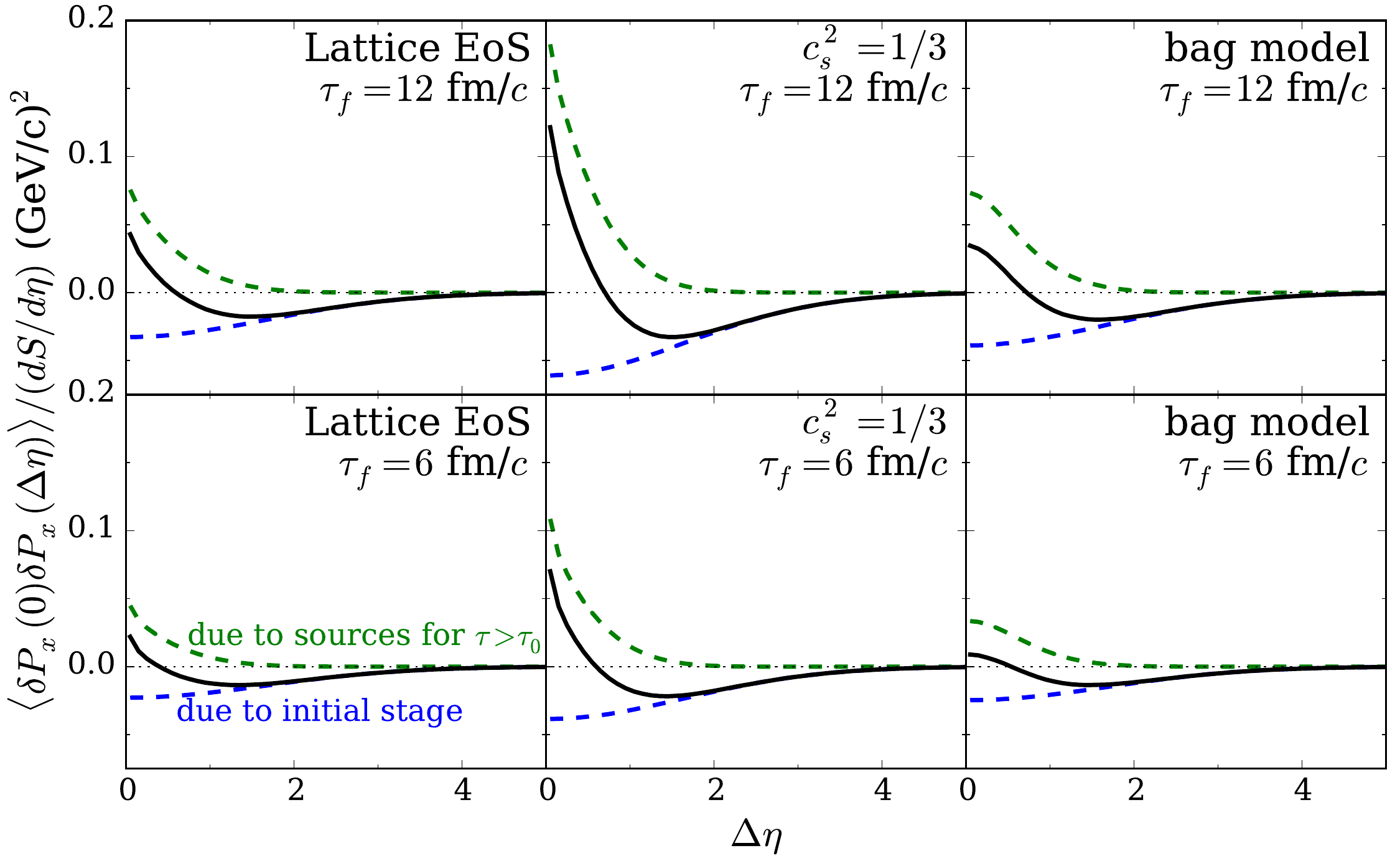}}
\caption{\label{fig:pxpxeos}(color online)
Transverse momentum fluctuations for three equations of state. Results from a realistic lattice equation of state, a stiff equation of state with $c_s^2=1/3$, and from a bag model equation of state with a first-order phase transition differ significantly. The stiffer equations of state reach higher initial temperatures, and given that all three have fixed final temperatures, have stronger sources which provide sharper maxima near $\Delta\eta=0$.}
\end{figure}

Both the source, which scales as $-TdT/d\tau$, and to a lesser degree the evolution, where the diffusion constant is $T\eta_s/s$, depend on the temperature and therefore on the equation of state. Figure \ref{fig:pxpxeos} demonstrates the sensitivity of momentum correlations to the equation of state by comparing results for three equations of state: an equation of state from lattice gauge theory, a much stiffer equation of state with $c^2=1/3$, and a bag model equation of state featuring a first-order phase transition. For the three calculations the same width, $\sigma_0=1$, is assumed for the initial correlation. By analyzing different centrality classes, one can gain some insight into the evolution of the equation of state with temperature. All calculations were performed assuming identical final temperatures of $T_f=155$ MeV, with the earlier entropy densities scaling as $s(\tau)=s(\tau_f)\tau_f/\tau$. By studying two values of $\tau_f$, 6 fm/$c$ an 12 fm/$c$, one can mimic the effect of considering two centralities, with the $\tau_f=12$ fm/$c$ representing the more central collisions. The more central collisions reach higher temperatures and thus their source functions generate stronger correlations. Stiffer equations of state also require higher initial temperatures, and thus have stronger correlations in the final state. Here the solid black lines represent the net correlation, the lower dashed (blue) lines show the correlation coming from the initial state, and the upper dashed lines (green) show the correlation coming from the source after $\tau_0$ when the matter thermalizes.

\section{Transverse Energy Fluctuations}
\label{sec:et}

As shown in Sec. \ref{sec:general} energy fluctuations are driven by changes to the specific heat, $T^2C_V$, whereas momentum fluctuations are related to changes in the inertial mass density, $(P+\epsilon)T$. Whereas charge and transverse momentum (transverse to the dimension of propagation) move diffusively, energy and longitudinal momentum move hydrodynamically, as sound waves. Fluctuations of the stress-energy tensor components $T_{00}$ and $T_{0z}$ cannot be treated independently because in hydrodynamics, one quantity drives the other. For non-zero net charge densities the two quantities also become intertwined with density fluctuations, and when transverse expansion is included the equations also include fluctuations of transverse momentum. For the illustrative purposes of this paper, we consider the idealized case with zero baryon density and no transverse expansion, where only the conserved quantities related to energy and longitudinal momentum must be simultaneously considered. 

This section is divided into several subsections. First, the response of the fluctuation of the energy density, $E$, and the longitudinal momentum density, $P_z$, to delta function perturbations is found analytically for the case of constant speed of sound with zero viscosity. In the second subsection, equations of motion are presented for the correlations $\langle E(0)E(\eta)\rangle$, $\langle E(0)P_z(\eta)\rangle$ and $\langle P_z(0)P_z(\eta)\rangle$. The equations are solved numerically and compared to the analytic results for the case of fixed speed of sound, and are solved for non-zero viscosity.

\subsection{Analytic results for constant speed of sound and zero viscosity}

Before deriving the analytic result for the propagation of a three-dimensional sound wave in Bjorken coordinates, it might be insightful to review the result for sound in a static, non-expanding Cartesian system. The solution for a spherical wave beginning at a point-like source at $x=y=z=t=0$ behaves like a pulse centered around $r=ct$, where $c$ is the speed of sound, with no energy or momentum deposited inside or outside the source. If one integrates the solution over $x$ and $y$, the paradoxical result is the solution behaves $\sim\delta(z\pm ct)$, with no energy or momentum excitation outside the pulse. The explanation for this result is provided in the appendix. For the more complicated case derived below, the similar integration over $x$ and $y$ also provides a delta function pulse originating from $\tau=\tau_0$ and $\eta=0$. The pulses move outward at the speed of sound relative to the static medium,  $\sim \delta(cs\pm\eta)$, where $s\equiv \ln(\tau/\tau_0$. However, for the expanding medium the pulses degrade over time, depositing energy and momentum into the medium between the pulses.

We consider the elements of the stress-energy tensor integrated over the transverse coordinates.
\begin{eqnarray}
\delta\tilde{T}_{\alpha\beta}&=&\int dxdy~\delta T_{\alpha\beta}(x,y,\eta,\tau).
\end{eqnarray}
Our goal is to find analytic solutions for the cases where $\delta\tilde{T}_{00}$ and $\delta\tilde{T}_{0z}$ begin life as delta function pulses at $\tau_0$.

Conservation of the stress-energy tensor, $\partial^\alpha T_{\alpha\beta}=0$, leads to the expressions.
\begin{eqnarray}
\partial_\tau (\tau \delta \tilde{T}_{00})&=&-\delta \tilde{T}_{zz}-\partial_\eta \delta \tilde{T}_{0z},\\
\nonumber
\partial_\tau (\tau \delta \tilde{T}_{0z})&=&-\delta \tilde{T}_{0z}-\partial_\eta \delta \tilde{T}_{zz}.
\end{eqnarray}
Using the definitions,
\begin{eqnarray}
E&\equiv&\tau\delta \tilde{T}_{00},\\
P_z&\equiv&\tau\delta\tilde{T}_{0z},\\
s&\equiv&\ln(\tau/\tau_0),
\end{eqnarray}
the equations of motion become
\begin{eqnarray}
\partial_sE&=&-c^2E-\partial_\eta P_z,\\
\nonumber
\partial_s P_z&=&-P_z-c^2\partial_\eta E,
\end{eqnarray}
where $c$ is the speed of sound. If one defines 
\begin{eqnarray}
\tilde{P}_z&\equiv&e^{(1+c^2)s/2}P/c,\\
\nonumber
\tilde{E}&\equiv&e^{(1+c^2)s/2}E,
\end{eqnarray}
the equations of motion can be expressed in a more symmetric fashion,
\begin{eqnarray}
\frac{\partial}{\partial cs}\tilde{E}&=&k\tilde{E}-\partial_\eta\tilde{P}_z,\\
\nonumber
\frac{\partial}{\partial cs}\tilde{P}_z&=&-k\tilde{P}_z-\partial_\eta\tilde{E},\\
\nonumber
k&\equiv&\frac{1-c^2}{2c}.
\end{eqnarray}
In particular, we are looking for solutions that at time $\tau=\tau_0$, or $s=0$, correspond to delta functions in either $\tilde{E}$ or $\tilde{P}_z$. The first solution, which matches the boundary condition $\tilde{E}(s=0,\eta)=\delta(\eta)$, $\tilde{P}_z(s=0,\eta)=0$, is given in terms of modified Bessel functions,
\begin{eqnarray}
\label{eq:EtildeExact}
\tilde{E}_a(cs,\eta)&=&\frac{1}{2}\delta(\eta-cs)+\frac{1}{2}\delta(\eta+cs)
+\frac{k}{2}I_0(kz)\Theta(cs-|\eta|)+\frac{kcs}{2z}I_1(kz)\Theta(cs-|\eta|),\\
\nonumber
\tilde{P}_{z,a}(cs,\eta)&=&\frac{1}{2}\delta(\eta-cs)-\frac{1}{2}\delta(\eta+cs)
+\frac{k\eta}{2z}I_1(kz)\Theta(cs-|\eta|),\\
\nonumber
z&\equiv&\sqrt{c^2s^2-\eta^2}.
\end{eqnarray}
Here, the subscript $a$ denotes that at $s=0$ the solution is a delta function in energy with no momentum. The net energy represented by this solution,
\begin{equation}
E_{\rm tot}=\int d\eta~(E\cosh\eta+cP_z\sinh\eta),
\end{equation}
is conserved. The terms $\cosh\eta$ and $\sinh\eta$ are the components of the relativistic four velocity, $\gamma$ and $\gamma v$, and account for the fact that $E$ and $P_z$ describe the energy and momentum in the frame of the observer moving with the fluid, whereas energy conservation requires expressing the energy is a fixed frame. The net momentum integrates to zero, because $E$ and $P_z$ are even and odd functions, respectively, of $\eta$.

This solution represents two delta function pulses receding from one another moving with velocities of $c$ if measured by an observer at the location of the pulse. The pulses decay with time, proportional to $e^{-(1+c^2)s/2}=(\tau/\tau_0)^{-(1+c^2)/2}$. In addition to the pulses, energy and momentum are deposited in the volume between the pulses as described by the Bessel functions.

A second solution, satisfying the boundary conditions $\tilde{E}(s=0,\eta)=0$, $\tilde{P}_z(s=0,\eta)=\delta(\eta)$, involves simply switching $\tilde{E}$ and $\tilde{P}_z$, and $k\rightarrow -k$, in the solution of Eq. (\ref{eq:EtildeExact}). 
\begin{eqnarray}
\label{eq:PtildeExact}
\tilde{P}_{z,b}(s,\eta)&=&\frac{1}{2}\delta(\eta-cs)+\frac{1}{2}\delta(\eta+cs)
-\frac{k}{2}I_0(kz)\Theta(cs-|\eta|)+\frac{kcs}{2z}I_1(kz)\Theta(cs-|\eta|),\\
\tilde{E}_{b}(s,\eta)&=&\frac{1}{2}\delta(cs-\eta)-\frac{1}{2}\delta(\eta+cs)
+\frac{k\eta}{2z}I_1(kz)\Theta(cs-|\eta|).
\end{eqnarray}
In this solution, a negative energy pulse moves to the left, while a positive energy pulse moves to the right. Both pulses have positive momentum.

If one has sources for energy and momentum, $S_E(\tau,\eta)$ and $S_P(\tau,\eta)$, the solutions in Eq.s (\ref{eq:EtildeExact}) and (\ref{eq:PtildeExact}), represent Green's functions for the respective sources,
\begin{eqnarray}
\label{eq:Ganal}
E(\tau,\eta)&=&\int_{\tau_0<\tau} d\tau_0 d\eta_0~\left(S_E(\tau_0,\eta_0)G_{EE}(\ln(\tau/\tau_0),\eta-\eta_0)
+S_P(\tau_0,\eta_0)G_{EP}(\ln(\tau/\tau_0),\eta-\eta_0)\right),\\
\nonumber
P(\tau,\eta)&=&\int_{\tau_0<\tau} d\tau_0 d\eta_0~\left(S_E(\tau_0,\eta_0)G_{PE}(\ln(\tau/\tau_0),\eta-\eta_0)
+S_P(\tau_0,\eta_0)G_{PP}(\ln(\tau/\tau_0),\eta-\eta_0)\right),\\
\nonumber
G_{EE}(s,\Delta\eta)&=&e^{-(1+c^2)s/2}\left\{
\frac{1}{2}\delta(\Delta\eta-cs)+\frac{1}{2}\delta(\Delta\eta+cs)
+\frac{k}{2}I_0(kz)\Theta(cs-|\Delta\eta|)+\frac{kcs}{2z}I_1(kz)\Theta(cs-|\Delta\eta|)
\right\},\\
\nonumber
G_{PE}(s,\Delta\eta)&=&c\cdot e^{-(1+c^2)s/2}\left\{
\frac{1}{2}\delta(\Delta\eta-cs)-\frac{1}{2}\delta(\Delta\eta+c)
+\frac{k\eta}{2z}I_1(kz)\Theta(cs-|\Delta\eta|)
\right\},\\
\nonumber
G_{EP}(s,\Delta\eta)&=&(1/c)e^{-(1+c^2)s/2}\left\{
\frac{1}{2}\delta(\Delta\eta-cs)-\frac{1}{2}\delta(\Delta\eta+c)
+\frac{k\eta}{2z}I_1(kz)\Theta(cs-|\Delta\eta|)
\right\},\\
\nonumber
G_{PP}(s,\Delta\eta)&=&e^{-(1+c^2)s/2}\left\{
\frac{1}{2}\delta(\Delta\eta-cs)+\frac{1}{2}\delta(\Delta\eta+cs)
-\frac{k}{2}I_0(kz)\Theta(cs-|\Delta\eta|)+\frac{kcs}{2z}I_1(kz)\Theta(cs-|\Delta\eta|)
\right\}.
\end{eqnarray}
In the limit of large initial times, $\tau_0\rightarrow\infty$, solutions are the same as those for a static, non-expanding medium, and the sound pulses no longer decay and only the delta functions remain of the Green's functions. In an expanding medium, $\tau_0<\infty$, the delta functions decay with time as $(\tau_0/\tau)^{1+c^2}$. The energy that disappears from the delta function then appears in between the delta functions as described above.

\subsection{Numerical solutions for a point source in Bjorken Coordinates}

Here, we consider the correlation from a point source in Bjorken coordinates with constant sound velocity, zero viscosity, and no transverse flow. We will compare two solutions, one where we convolute the energy and momentum from the analytic solutions for two receding waves described by Eq. (\ref{eq:Ganal}), and secondly for numerical solutions to the equations for the correlations of transverse energy and longitudinal momentum according to the paradigm laid out in Sec. \ref{sec:general}. The equations for this latter solution are complicated by the fact that transverse energy and longitudinal momentum are inextricably linked by the equations of hydrodynamics.

Our goal here is to find equations of conservation for energy and momentum expressed in the format that $\partial_t\rho=-\nabla\cdot({\rm something})$, where $\rho$ denotes the components to the stress energy tensor $\delta \tilde{T}_{00}$ and $\delta \tilde{T}_{0z}$. To simplify the problem these quantities are integrated over $x$ and $y$ for a given $\eta$, and effectively ignore the transverse coordinates.
\begin{equation}
\label{eq:Ttildedef}
\tilde{T}_{\alpha\beta}=\frac{1}{A}\int dxdy~T_{\alpha\beta},
\end{equation}
where $A$ is the transverse area. We consider these elements in the longitudinally co-moving Bjorken frame. The equations of motion for the stress-energy tensor are
\begin{eqnarray}
\label{eq:setensor}
\partial_\tau \delta \tilde{T}_{00}&=&-\frac{1}{\tau}(\delta \tilde{T}_{00}+\delta \tilde{T}_{zz})-\frac{1}{\tau}\partial_\eta \delta \tilde{T}_{0z},\\
\partial_\tau \delta \tilde{T}_{0z}&=&-\frac{2}{\tau}\delta \tilde{T}_{0z}-\frac{1}{\tau}\partial_\eta \delta \tilde{T}_{zz}.
\end{eqnarray}
It is is easy to take the second equation and consider $\tau^2 \delta \tilde{T}_{0z}$,
\begin{equation}
\partial_\tau \left(\tau^2\delta \tilde{T}_{0z}\right)=-\partial_\eta (\tau \delta \tilde{T}_{zz}). 
\end{equation}
One can show that the correlation, 
\begin{equation}
C_{LL}(\tau,\eta)\equiv\tau^4\langle \delta \tilde{T}_{0z}(\tau,0)\delta \tilde{T}_{0z}(\tau,\eta)\rangle,
\end{equation}
evolves according to
\begin{eqnarray}
\partial_\tau C_{LL}&=&-2\tau^3\partial_\eta\langle \delta \tilde{T}_{0z}(\tau,0)\delta \tilde{T}_{zz}(\tau,\eta)\rangle.
\end{eqnarray}
The integral $\int d\eta ~C_{LL}(\eta)$ then integrates to a constant. The first two factors of $\tau$ account for switching to momentum per unit rapidity, rather than that per unit length. The last two factors account for the fact that asymptotically $\tau \delta \tilde{T}_{0z}$ falls proportionally to $1/\tau$ due to the velocity of any particles approaching the velocity of the co-moving matter. Thus $\delta \tilde{T}_{0z}$ effectively vanishes as $\tau\rightarrow\infty$, and this correlation is not useful in the context of heavy-ion collisions. This was expected because asymptotically particles have only transverse momentum or transverse energy when observed in the co-moving frame. One can measure $E_tE_t$ correlations as a function of relative rapidity but not $p_zp_z$ correlations because their velocities are the very quantities used to determine rapidity.

Our next goal is to find a good measure of a density-like operator that effectively represents the transverse energy while maintaining local conservation, $\partial_t\rho=\partial_\eta({\rm something})$. To do this we return to the equations of motion of the stress-energy tensor, Eq. (\ref{eq:setensor}). We consider equations for quantities multiplied by combinations of $\sinh\eta$ and $\cosh\eta$.
\begin{eqnarray}
\partial_\tau (\tau \delta \tilde{T}_{00})&=&-\delta \tilde{T}_{zz}-\partial_\eta \delta \tilde{T}_{0z},\\
\partial_\tau (\tau \delta \tilde{T}_{0z})&=&-\delta \tilde{T}_{0z}-\partial_\eta \delta \tilde{T}_{zz},\\
\partial_\tau (\tau \delta \tilde{T}_{00}\cosh\eta)&=&-\delta \tilde{T}_{zz}\cosh\eta-\partial_\eta(\delta \tilde{T}_{0z}\cosh\eta)+\delta \tilde{T}_{0z}\sinh\eta,\\
\partial_\tau (\tau \delta \tilde{T}_{0z}\sinh\eta)&=&-\delta \tilde{T}_{0z}\sinh\eta-\partial_\eta(\delta \tilde{T}_{zz}\sinh\eta)+\delta \tilde{T}_{zz}\cosh\eta.
\end{eqnarray}
Adding the two equations brings one to the desired form. Similarly, one could have switched $\cosh\eta$ and $\sinh\eta$ and also obtained the desired form,
\begin{eqnarray}
\partial_\tau(\tau \delta \tilde{T}_{00}\cosh\eta+\tau \delta \tilde{T}_{0z}\sinh\eta)&=&-\partial_\eta\left(\delta \tilde{T}_{0z}\cosh\eta+\delta \tilde{T}_{zz}\sinh\eta\right),\\
\partial_\tau(\tau \delta \tilde{T}_{00}\sinh\eta+\tau \delta \tilde{T}_{0z}\cosh\eta)&=&-\partial_\eta\left(\delta \tilde{T}_{0z}\sinh\eta+\delta \tilde{T}_{zz}\cosh\eta\right),
\end{eqnarray}
One might consider the correlation
\begin{equation}
C_{00}(\tau,\eta_1,\eta_2)\equiv\tau^2\langle
 (\tau \delta \tilde{T}_{00}(\tau,\eta_1)\cosh\eta_1+\tau \delta \tilde{T}_{0z}(\tau,\eta_1)\sinh\eta_1)
 (\tau \delta \tilde{T}_{00}(\tau,\eta_2)\cosh\eta_2+\tau \delta \tilde{T}_{0z}(\tau,\eta_2)\sinh\eta_2)\rangle,
\end{equation}
but the correlation would not be merely a function of relative $\eta$ and the equations for total charge conservation, integrated over $\eta$ would explode at the far away limits. A solution to this is to take difference of this correlation with one switching $\cosh\eta$ and $\sinh\eta$,
\begin{eqnarray}
C_{TT}(\tau,\eta_1,\eta_2)&\equiv&\tau^2\langle
 (\tau \delta \tilde{T}_{00}(\tau,\eta_1)\cosh\eta_1+\tau \delta \tilde{T}_{0z}(\tau,\eta_1)\sinh\eta_1)
 (\tau \delta \tilde{T}_{00}(\tau,\eta_2)\cosh\eta_2+\tau \delta \tilde{T}_{0z}(\tau,\eta_2)\sinh\eta_2)\rangle\\
 &-&\tau^2\langle
 (\tau \delta \tilde{T}_{00}(\tau,\eta_1)\sinh\eta_1+\tau \delta \tilde{T}_{0z}(\tau,\eta_1)\cosh\eta_1)
 (\tau \delta \tilde{T}_{00}(\tau,\eta_2)\sinh\eta_2+\tau \delta \tilde{T}_{0z}(\tau,\eta_2)\cosh\eta_2)\rangle~.
\end{eqnarray}
This simplifies to
\begin{equation}
C_{TT}(\tau,\eta=\eta_2-\eta_1)=
\tau^2\cosh\eta\langle \delta \tilde{T}_{00}(\tau,0)\delta \tilde{T}_{00}(\tau,\eta)-\delta \tilde{T}_{0z}(\tau,0)\delta \tilde{T}_{0z}(\tau,\eta)\rangle
+2\tau^2\sinh\eta\langle \delta \tilde{T}_{00}(\tau,0)\delta \tilde{T}_{0z}(\tau,\eta)\rangle.
\end{equation}
Again, this derivation used the fact that $\langle \delta \tilde{T}_{00}(\tau,0)\delta \tilde{T}_{0z}(\tau,\eta)\rangle$ is an odd function of $\eta$.
The correlator obeys the equations of motion
\begin{eqnarray}
\partial_\tau C_{TT}(\tau,\eta=\eta_2-\eta_1)&=&
-2\partial_\eta\tau\left\{
\cosh\eta\langle\delta \tilde{T}_{00}(\tau,0)\delta \tilde{T}_{0z}(\tau,\eta)
-\delta \tilde{T}_{0z}(\tau,0)\delta \tilde{T}_{zz}(\tau,\eta)\rangle\right.\\
\nonumber
&&\hspace*{40pt}\left.
+\sinh\eta\langle\delta \tilde{T}_{00}(\tau,0)\delta \tilde{T}_{zz}(\tau,\eta)-\delta \tilde{T}_{0z}(\tau,0)\delta \tilde{T}_{0z}(\tau,\eta)
\rangle\right\}
\end{eqnarray}

Rather than considering the correlations between energy-like and energy-like or between momentum-like and momentum-like fluctuations, one could consider the correlation between the energy and momentum like correlations,
\begin{eqnarray}
\nonumber
C_{LT}(\tau,\eta_1,\eta_2)&\equiv&\tau^2\langle
 (\tau \delta \tilde{T}_{00}(\tau,\eta_1)\cosh\eta_1+\tau \delta \tilde{T}_{0z}(\tau,\eta_1)\sinh\eta_1)
 (\tau \delta \tilde{T}_{00}(\tau,\eta_2)\sinh\eta_2+\tau \delta \tilde{T}_{0z}(\tau,\eta_2)\cosh\eta_2)\rangle\\
 \nonumber
 &-&\tau^2\langle
 (\tau \delta \tilde{T}_{00}(\tau,\eta_1)\sinh\eta_1+\tau \delta \tilde{T}_{0z}(\tau,\eta_1)\cosh\eta_1)
 (\tau \delta \tilde{T}_{00}(\tau,\eta_2)\cosh\eta_2+\tau \delta \tilde{T}_{0z}(\tau,\eta_2)\sinh\eta_2)\rangle\\
&=&\tau^2\langle \delta \tilde{T}_{00}(\tau,0)\delta \tilde{T}_{00}(\tau,\eta)
 -\delta \tilde{T}_{0z}(\tau,0)\delta \tilde{T}_{0z}(\tau,\eta)\rangle\sinh\eta
 +2\tau^2\langle \delta \tilde{T}_{00}(\tau,0)\delta \tilde{T}_{0z}(\tau,\eta)\rangle\cosh\eta
\end{eqnarray}
The equations of motion then become
\begin{eqnarray}
\partial_\tau C_{LT}(\tau,\eta)&=&
-2\partial_\eta\tau\left\{
\cosh\eta\langle\delta \tilde{T}_{00}(\tau,0)\delta \tilde{T}_{zz}(\tau,\eta)-\delta \tilde{T}_{0z}(\tau,0)\delta \tilde{T}_{0z}(\tau,\eta)\rangle\right.\\
\nonumber
&&\left.+\sinh\eta\langle\delta \tilde{T}_{00}(\tau,0)\delta \tilde{T}_{0z}(\tau,\eta)+\delta \tilde{T}_{zz}(\tau,0)\delta \tilde{T}_{0z}(\tau,\eta)\rangle\right\}~.
\end{eqnarray}

Summarizing, the three correlators for transverse energy and longitudinal momentum are defined here as:
\begin{eqnarray}
C_{TT}(\tau,\eta)&=&
\tau^2\langle \delta \tilde{T}_{00}(\tau,0)\delta \tilde{T}_{00}(\tau,\eta)-\delta \tilde{T}_{0z}(\tau,0)\delta \tilde{T}_{0z}(\tau,\eta)\rangle\cosh\eta
+2\tau^2\langle \delta \tilde{T}_{00}(\tau,0)\delta \tilde{T}_{0z}(\tau,\eta)\rangle\sinh\eta\\
\nonumber
C_{LT}(\tau,\eta)&=&\tau^2\langle \delta \tilde{T}_{00}(\tau,0)\delta \tilde{T}_{00}(\tau,\eta)
 -\delta \tilde{T}_{0z}(\tau,0)\delta \tilde{T}_{0z}(\tau,\eta)\rangle\sinh\eta
 +2\tau^2\langle \delta \tilde{T}_{00}(\tau,0)\delta \tilde{T}_{0z}(\tau,\eta)\rangle\cosh\eta\\
\nonumber
C_{LL}(\tau,\eta)&=&\tau^4\langle \delta \tilde{T}_{0z}(\tau,0)\delta \tilde{T}_{0z}(\tau,\eta)\rangle.
\end{eqnarray}
At large times only $C_{TT}$ survives because for each particle the rapidity $y\rightarrow\eta$ at large times when the temperature vanishes and $\tilde{T}_{0z}$ vanishes. This correlation then represents the measurable correlation of transverse energy,
\begin{equation}
\label{eq:CTTinfty}
C_{TT}(\tau\rightarrow\infty,\Delta y)=\left\langle\delta\frac{dE_T}{d\eta}(0)\delta\frac{dE_T}{dy}(\Delta y)\right\rangle \cosh \Delta y.
\end{equation}
The three equations of motion are:
\begin{eqnarray}
\label{eq:summary1}
\partial_\tau C_{TT}(\tau,\eta=\eta_2-\eta_1)&=&
-2\partial_\eta\tau\left\{
\cosh\eta\langle\delta \tilde{T}_{00}(\tau,0)\delta \tilde{T}_{0z}(\tau,\eta)
+\delta \tilde{T}_{zz}(\tau,0)\delta \tilde{T}_{0z}(\tau,\eta)\rangle\right.\\
\nonumber
&&\hspace*{40pt}\left.
+\sinh\eta\langle\delta \tilde{T}_{00}(\tau,0)\delta \tilde{T}_{zz}(\tau,\eta)-\delta \tilde{T}_{0z}(\tau,0)\delta \tilde{T}_{0z}(\tau,\eta)
\rangle\right\}\\
\nonumber
\partial_\tau C_{LT}(\tau,\eta)&=&
-2\partial_\eta\tau\left\{
\cosh\eta\langle\delta \tilde{T}_{00}(\tau,0)\delta \tilde{T}_{zz}(\tau,\eta)-\delta \tilde{T}_{0z}(\tau,0)\delta \tilde{T}_{0z}(\tau,\eta)\rangle\right.\\
\nonumber
&&\hspace*{40pt}\left.+\sinh\eta\langle\delta \tilde{T}_{00}(\tau,0)\delta \tilde{T}_{0z}(\tau,\eta)+\delta \tilde{T}_{zz}(\tau,0)\delta \tilde{T}_{0z}(\tau,\eta)\rangle\right\}\\
\nonumber
\partial_\tau C_{LL}&=&-2\tau^3\partial_\eta\langle \delta \tilde{T}_{0z}(\tau,0)\delta \tilde{T}_{zz}(\tau,\eta)\rangle.
\end{eqnarray}
The right-hand-side of each equation of motion is of the form $\partial_\mu(\cdots)$, therefore in the absence of sources each correlation integrates to a constant. Further, as $\eta\rightarrow 0$, the correlations $C_{TT}$ and $C_{LL}$ can be identified with susceptibilities, while $C_{LT}(\eta\rightarrow 0)=0$. The form then fits the paradigm discussed in Sec. \ref{sec:general} and the source functions can be identified with changing susceptibilities.

The correlations in the right-hand side of Eq. (\ref{eq:summary1}) can be expressed in terms of $C_{TT}$, $C_{LT}$ and $C_{LL}$, which will then make it possible to derive a closed set of equations.
\begin{eqnarray}
\tau^2\langle\delta \tilde{T}_{00}(\tau,0)\delta \tilde{T}_{00}(\tau,\eta)\rangle&=&
C_{TT}(\tau,\eta)\cosh\eta -C_{LT}(\tau,\eta)\sinh\eta +C_{LL}(\tau,\eta)/\tau^2,\\
\nonumber
\tau^2\langle\delta \tilde{T}_{0z}(\tau,0)\delta \tilde{T}_{0z}(\tau,\eta)\rangle &=& C_{LL}(\tau,\eta)/\tau^2,\\
\nonumber
\tau^2\langle\delta \tilde{T}_{00}(\tau,0)\delta \tilde{T}_{0z}(\tau,\eta)\rangle &=&
(C_{LT}(\tau,\eta)\cosh\eta -C_{TT}(\tau,\eta)\sinh\eta)/2.
\end{eqnarray}
If one assumes perfect hydrodynamics and ignores transverse flow, $\delta \tilde{T}_{zz}=c^2\delta \tilde{T}_{00}$, the three equations of motion can be written self-consistently purely in terms of the three correlators. This gives the closed set of equations,
\begin{eqnarray}
\label{eq:idealdCdt}
\nonumber
\tau\partial_\tau C'_{TT}+2\partial_\eta\bigg\{
\frac{(1+c^2)}{2}\cosh\eta(\cosh\eta~C_{LT}-\sinh\eta~C_{TT})
+c^2\sinh\eta(\cosh\eta~C_{TT}-\sinh\eta~C_{LT}&+&(1/\tau^2)C_{LL})\\
-\sinh\eta(1/\tau^2)C_{LL}\bigg\}&=&(S_{TT}/A)\delta(\eta),\\
\tau\partial_\tau C'_{LT}+2\partial_\eta\bigg\{
c^2\cosh\eta(\cosh\eta~C'_{TT}-\sinh\eta~C_{LT}+(1/\tau^2)C'_{LL})
-\cosh\eta~(1/\tau^2)C'_{LL}&&\\
\nonumber
+\frac{(1+c^2)}{2}\sinh\eta(\cosh\eta~C'_{LT}-\sinh\eta~C'_{TT})\bigg\}&=&(S_{LT}/A)\delta(\eta),\\
\nonumber
\partial_\tau C'_{LL}+\tau^2c^2\partial_\eta\left\{
-\cosh\eta~C'_{LT}+\sinh\eta~C'_{TT}
\right\}&=&(S_{LL}/A)\delta(\eta).
\end{eqnarray}
Here, the prime again denotes that one is neglecting the part of the correlation at $\eta=0$. The cross-sectional area is $A$, and the sources and correlations all have strengths inverse to the area due to the way in which $\tilde{T}$ was defined in Eq. (\ref{eq:Ttildedef}).

The source functions in Eq.s (\ref{eq:idealdCdt}) can be found by considering the rate of change of the equilibrated susceptibilities.
\begin{eqnarray}
\chi_{LL}&=&
\tau\int d^3r~\tau^2\langle \tilde{T}_{0z}(0,0)\tilde{T}_{0z}(0,\vec{r})\rangle_{\rm eq}\\
\nonumber
&=&\tau^2 (P+\epsilon)T,\\
\nonumber
\chi_{TT}&=&\int d^3r~ \left\{\langle \tilde{T}_{00}(0,0)\tilde{T}_{00}(0,\vec{r})\rangle_{\rm eq}
-\langle \tilde{T}_{0z}(0,0)\tilde{T}_{0z}(0,\eta)\rangle_{\rm eq}\right\}\\
\nonumber
&=&(P+\epsilon)T\left(\frac{1}{c^2}-1\right),\\
\nonumber
\chi_{LT}&=&0.
\end{eqnarray}
As discussed in Sec. \ref{sec:general}, if the system is neutral $s=(P+\epsilon)/T$ and the source functions are then $-s\tau \partial_\tau(\chi/s)$,
\begin{eqnarray}
\label{eq:SET}
S_{TT}(\tau)&=&-\frac{dS}{d\eta}\partial_\tau\left\{T^2\left(\frac{1}{c^2}-1\right)\right\}\\
\nonumber
S_{LL}(\tau)&=&-\frac{dS}{d\eta}\partial_\tau\left\{\tau^2 T^2\right\},\\
\nonumber
S_{LT}&=&0,
\end{eqnarray}
where $dS/d\eta$ is the entropy per unit rapidity. These expressions ignore transverse expansion. Transverse expansion keeps one from identifying the susceptibility in such a straight-forward manner. For example, once transverse motion is considered $\delta\tilde{T}_{00}$ can no longer be equated with $\delta\epsilon$.

\begin{figure}
\centerline{\includegraphics[width=0.6\textwidth]{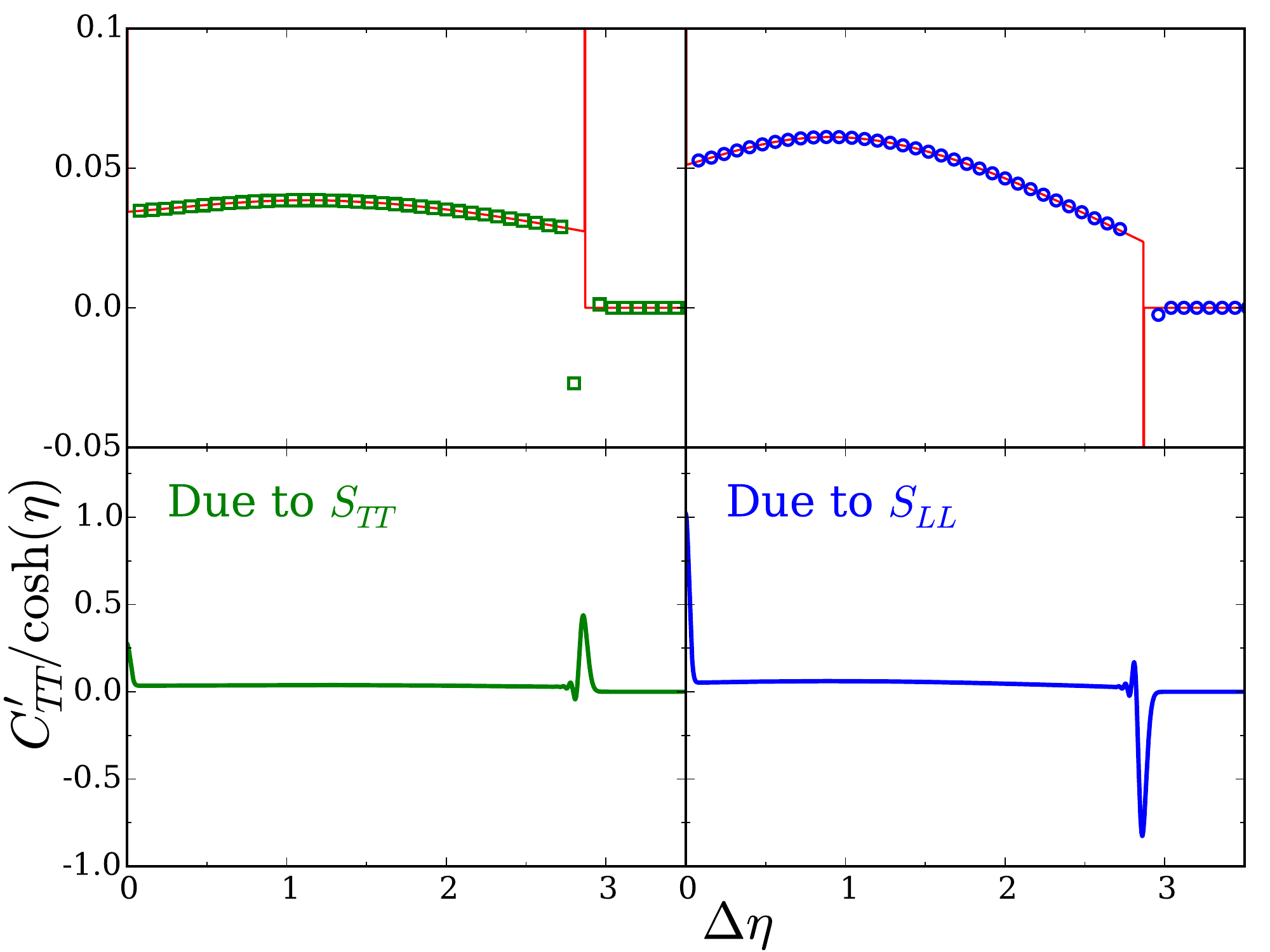}}
\caption{\label{fig:anal}(color online)
The transverse energy correlator, $C'_{TT}$ at $\tau=12$ fm/$c$ due to point-like sources at $\tau_0=1$ fm/$c$ and propagating due to ideal hydrodynamics. Transverse expansion is neglected. The left-side panels show the correlation due to a source $S_{TT}$, with a strength that integrates to unity, whereas the right-side panels present the correlation from a source $S_{LL}$, also with unit strength. The upper/lower panels are the same results, but in the upper panels the scale is decreased and the numerical solutions are represented by symbols. Numerical calculations were performed using the explicit equations of motion of the correlator from Eq. (\ref{eq:idealdCdt}) and are represented by the green squares or blue circles in the upper panels and by lines in the lower panels. The correlator was also calculated from the explicit analytic solutions of the Green functions, which are then convoluted with one another according to Eq. (\ref{eq:Gconvolute}) and are represented by red lines in the upper panels. For the numerical solutions the source function pulse is represented by a finite-width Gaussian, which leads to a small discrepancy with the analytic method which assumes delta function sources. The sharp peaks come from the contribution where $\eta_1$ and $\eta_2$ come from the forward and/or backward-moving pulses.}
\end{figure}

Solutions to Eq.s (\ref{eq:idealdCdt}) are shown in Fig. \ref{fig:anal} for the case of two point like sources and assuming a simple equation of state, $P=\epsilon/3$. The left-side panels present $C'_{TT}$ at $\tau=12$ fm/$c$ for a point-like source at $\tau_0=1$ fm/$c$. 
\begin{eqnarray}
\label{eq:STTpoint}
S_{TT}(\tau)&=&\delta(\tau-\tau_0)/A.
\end{eqnarray}
With this form, the correlation integrates to unity for all times $\tau>\tau_0$,
\begin{equation}
\int_{-\infty}^\infty d\eta C_{TT}(\tau,\eta)=\int_{-\infty}^\infty d\tau S_{TT}(\tau)=1.
\end{equation}
Solutions are compared to the analytic results of the previous subsections. However, the analytic results include delta functions at $\eta=0$ and $\eta=2c\ln(\tau/\tau_0)$. These correspond to the contributions from the two receding plane waves which were generated by the source. Of course, delta functions cannot be implemented numerically. Instead, the delta functions were generated by Gaussian forms with a finite width of 0.02 in $\eta$. This led to finite-sized peaks where the delta functions were located, plus some small oscillations near the delta functions. The numerical solutions indeed integrated to unity as demanded by the conservation laws. 

The second source that was considered was
\begin{eqnarray}
\label{eq:SLLpoint}
S_{LL}&=&\delta(\eta)\delta(\tau-\tau_0)/A,
\end{eqnarray}
again with $c^2$ fixed at $1/3$ and unit area $A$. For this source the contribution to $G_{TT}$ integrates to zero, even though that contribution is even in $\eta$ because $S_{TT}=0$.

Comparisons to the analytic solutions in Eq. (\ref{eq:Ganal}) are also displayed in Fig. \ref{fig:anal}. In that case the correlation corresponding to the transverse energy correlations $C_{TT}(\eta)$ is found by convoluting over the Green's functions,
\begin{eqnarray}
\label{eq:Gconvolute}
\left.C'_{TT}(\eta)\right|_{{\rm due~to~}S_{TT}}&=&\int d\eta_1d\eta_2~\left\{
G_{EE}(\eta_1)G_{EE}(\eta_2)\cosh(\eta)
-G_{PE}(\eta_1)G_{PE}(\eta_2)\cosh(\eta)\right.\\
\nonumber
&&\left.\hspace*{40pt} +2G_{EE}(\eta_1)G_{PE}(\eta_2)\sinh(\eta)
\right\}\delta(\eta-(\eta_2-\eta_1)).\\
\nonumber
\left.C'_{TT}(\eta)\right|_{{\rm due~to~}S_{LL}}&=&\left.C'_{TT}(\eta)\right|_{{\rm due~to~}S_{TT}}
+\int d\eta_1d\eta_2~\left\{
G_{EP}(\eta_1)G_{EP}(\eta_2)\sinh(\eta)
-G_{PP}(\eta_1)G_{PP}(\eta_2)\sinh(\eta)\right.\\
\nonumber
&&\left.\hspace*{40pt} +2G_{EP}(\eta_1)G_{PP}(\eta_2)\cosh(\eta)
\right\}\delta(\eta-(\eta_2-\eta_1)).
\end{eqnarray}
Although the Green's functions were analytic, the convolution was performed numerically. As one can see there is a peak at zero relative spatial rapidity $\eta$. This peak comes from $\eta_1$ and $\eta_2$ both coming from the same forward going, or both coming from the same backward going, delta function pulses. This is qualitatively different behavior than in the charge correlations, where the separation is diffusive, due to the fact that for hydrodynamics a finite amount of the propagation from a pulse is contained in delta functions. Thus, even though the energy moves from its original location via forward and backward pulses, one has a finite probability of having either two particles from the same pulse, or one from one pulse and one from the other. The correlators, $C'_{TT}$, were scaled by $\cosh\eta$ so that the correlators would better represent the correlations in transverse energy observed asymptotically described by Eq. (\ref{eq:CTTinfty}).

\subsubsection{Adding Viscosity}

Here, the work in the previous section is extended to incorporate viscosity. Transverse expansion remains ignored. According to the Navier-Stokes equations the contribution to the stress-energy tensor from shear is,
\begin{eqnarray}
\delta T_{zz}^{\rm(shear)}=-\frac{4\delta\eta_s}{3\tau}-\frac{4\eta_s}{3\tau}\partial_\eta\delta v_z.
\end{eqnarray}
The fluctuation of the shear viscosity, $\delta\eta_s$, comes from fluctuations of the density, whereas the fluctuation of the collective velocity, $\delta v_z$, is related to fluctuations of the momentum density,
\begin{eqnarray}
\delta T_{0z}&=&\delta \left\{\left(P+\epsilon-\frac{4\eta_s}{3\tau}\right)v_z\right\},\\
\nonumber
\delta v_z&=&\frac{\delta T_{0z}}{P+\epsilon-\frac{4\eta_s}{3\tau}}.
\end{eqnarray}
This then gives
\begin{eqnarray}
\delta T_{zz}^{\rm(shear)}&=&\frac{4}{3\tau}\frac{d\eta_s}{d\epsilon}\delta T_{00}
-\frac{4\eta_s/(3\tau)}{P+\epsilon-4\eta_s/(3\tau)}
\partial_\eta \delta T_{0z},\\
\nonumber
\delta T_{zz}&=&c^{\prime 2}\delta T_{00}-\chi \partial_\eta\delta T_{0z},\\
c^{\prime 2}&\equiv& c^2-\frac{4\eta_s}{3\tau},\\
\nonumber
\chi&\equiv&\frac{4\eta_s/(3\tau)}{P+\epsilon-4\eta_s/(3\tau)}.
\end{eqnarray}
The Equations of motion in Eq.s(\ref{eq:summary1}) can be recalculated with these modifications, which in matrix form  become
\begin{equation}
C'=\left(\begin{array}{c}
C'_{TT}\\C'_{LT}\\C'_{LL}\end{array}\right),~~~S=\left(\begin{array}{c}
S_{TT}\\S_{LT}\\S_{LL}\end{array}\right)
\end{equation}
\begin{eqnarray}
\label{eq:masteretet}
\tau \partial_\tau C'+AC' +B\partial_\eta C' +D\partial^2_\eta C'&=&S,\\
\nonumber
A&=&\left(\begin{array}{ccc}
-2\kappa(\cosh^2\eta+\sinh^2\eta) & 4\kappa\cosh\eta\sinh\eta & -4\kappa\cosh\eta/\tau^2\\
-4\kappa\cosh\eta\sinh\eta & 2\kappa(\cosh^2\eta+\sinh^2\eta) & -4\kappa\sinh\eta/\tau^2\\
c_s^{\prime 2}\cosh\eta & -c_s^{\prime 2}\sinh\eta & 0
\end{array}\right)\\
\nonumber && +
\chi\left(\begin{array}{ccc}
(\cosh\eta^2+\sinh^2\eta) & -2\cosh\eta\sinh\eta & 0\\
2\cosh\eta\sinh\eta & -(\cosh^2\eta+\sinh^2\eta) & 0\\
0 & 0 & 0
\end{array}\right)\\
\nonumber
B&=&\left(\begin{array}{ccc}
-2\kappa\sinh\eta\cosh\eta & 2c_s^{\prime 2}+2\kappa\cosh^2\eta & -4\kappa\sinh\eta/\tau^2\\
2c_s^{\prime 2}-2\kappa \sinh^2\eta & 2\kappa\sinh\eta\cosh\eta & -4\kappa\cosh\eta/\tau^2\\
c_s^{\prime 2}\sinh\eta & -c_s^{\prime 2}\cosh\eta & 0
\end{array}\right)\\
\nonumber
&&+\chi\left(\begin{array}{ccc}
3\sinh\eta\cosh\eta & -(2\sinh^2\eta+\cosh^2\eta) & 2\sinh\eta/\tau^2\\
(2\cosh^2\eta+\sinh^2\eta) & -3\cosh\eta\sinh\eta & 2\cosh\eta/\tau^2\\
0 & 0 & 0\end{array}\right)\\
\nonumber
D&=&\chi\left(\begin{array}{ccc}
\sinh^2\eta & -\cosh\eta\sinh\eta & 2\cosh\eta/\tau^2\\
\cosh\eta\sinh\eta & -\cosh^2\eta & 2\sinh\eta/\tau^2\\
0 & 0 & -2
\end{array}\right)
\end{eqnarray}
Here $\kappa\equiv (1-c_s^{\prime 2})/2$.

\begin{figure}
\centerline{\includegraphics[width=0.45\textwidth]{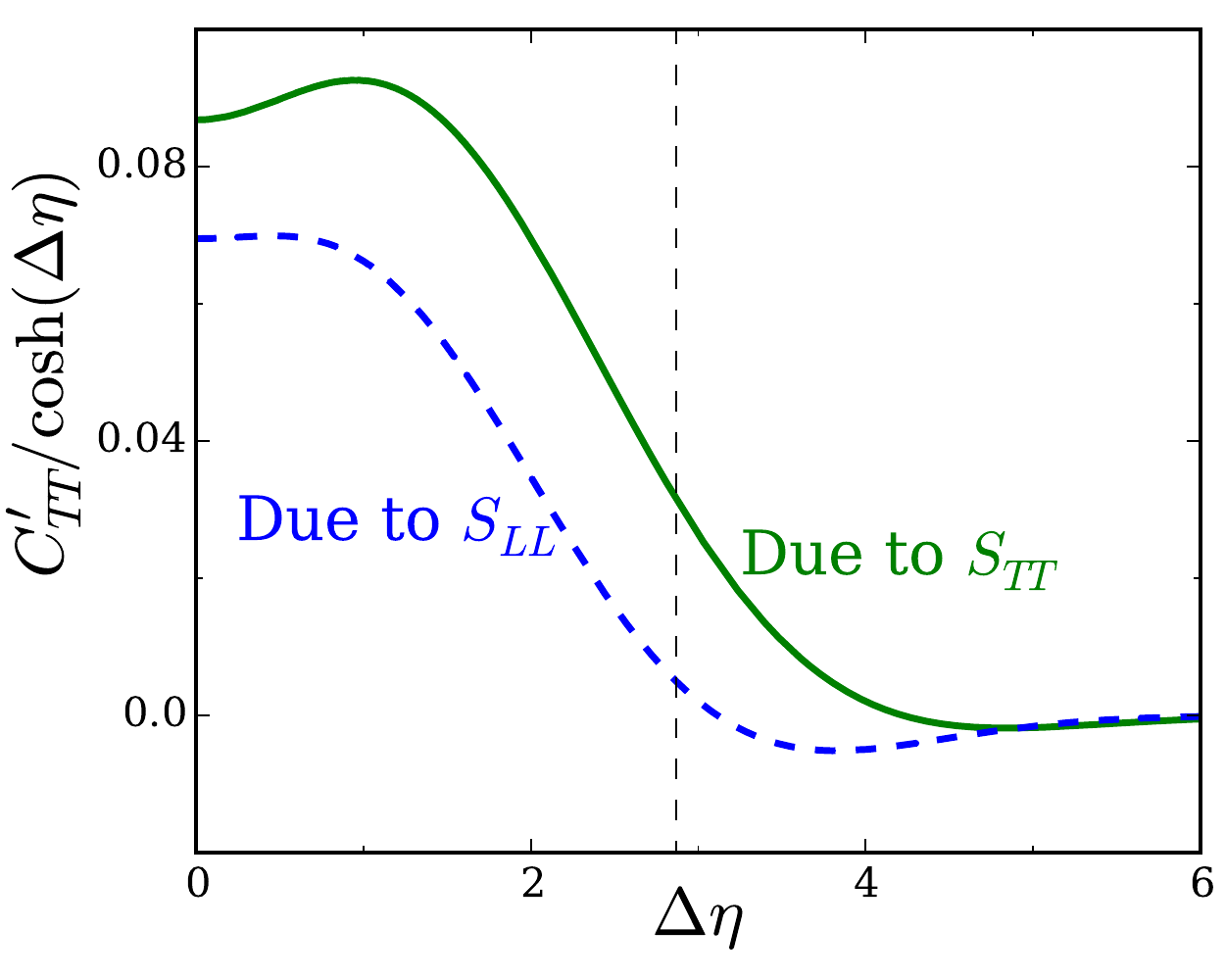}}
\caption{\label{fig:visc}(color online) 
Correlations for the same conditions as in Fig. \ref{fig:anal} were calculated with the effects of viscosity. The small viscosity, $\eta_s=s/4\pi$, spread contribution from the delta function pulses, which for the previous calculation had appeared at $\eta=0$ and $\eta=2c\ln(\tau/\tau_0)$, noted by the thin vertical dashed line. The upper line (solid green) shows the correlation of the transverse energy due to point-like source for the transverse energy, $S_{TT}$, at $\tau_0$, whereas the lower line (dashed blue) provides the correlation from a point-like source of the longitudinal momentum. }
\end{figure}

These equations were solved numerically, for the same circumstances as the non-viscous solutions displayed in Fig. (\ref{fig:anal}): a source function at $\tau=1$ fm/$c$ that propagates according to a constant speed of sound until $\tau=12$ fm/$c$. The resulting correlators $C_{TT}(\eta)$ are shown in Fig. \ref{fig:visc} for the same point-like sources $S_{TT}$ and $S_{LL}$ from Eq.s (\ref{eq:STTpoint}) and (\ref{eq:SLLpoint}). The one difference between these solutions and those presented in Fig. \ref{fig:anal} is that a small shear viscosity, $\eta=s/4\pi$, was assumed. The resulting correlation was very significantly broadened by the viscosity and the delta function contributions were no longer visible. 

Once viscosity is incorporated into the solution, the spread of the transverse energy correlator becomes somewhat more Gaussian in nature but still has distinct non-Gaussian features. At small $\eta$ correlations might even rise modestly with $\eta$, and at large $\eta$ the functions dip below zero. The shapes are driven by characteristic widths set by the distance the pulses, in ideal hydrodynamics would travel, $\eta_{\rm char}\sim 2c\ln(\tau/\tau_0)$. For $\tau/\tau_0=12$ fm/$c$, these widths are 2.86 units of relative rapidity if the speed of sound is $c^2=1/3$. For contributions from the source function at later times, the delta function features will have had less time to dissipate, and for softer equations of state the width will be spread less due to the reduced values for the speed of sound.
 
\subsection{Continuous Source Functions and Sensitivity to the Equation of State}

Ultimately, the measured transverse energy correlations are the integrated sum of contributions from the source functions at each point in space, as listed in Table \ref{table:chiab}. For the  boost-invariant case with no transverse expansion the source functions are purely functions of $\tau$. The source functions relevant for transverse energy, $S_{TT}$ and $S_{LL}$ given in Eq.s (\ref{eq:SET}), are simple functions of the temperature if given the equation of state. If matter equilibrates to the point that one understands the stress-energy tensor and its fluctuations at time $\tau_0$, the source functions at later times can be reasonably well calculated. However, understanding the sources at earlier times, or understanding the correlations $C'_{TT}$, $C'_{LL}$ and $C'_{LT}$ at $\tau_0$, is challenging. The only constraints are that $-C'_{TT}$ and $-C'_{LL}$ integrate to their respective susceptibilities, and that $C'_{LT}$ is an odd function in $\eta=\eta_2-\eta_1$, and should thus integrate to zero. If one foregoes the assumption of boost invariance, even these assumptions need to be altered. The structures are affected by details of energy deposition at early times and by the mechanism of jet equilibration. However, from the results of the previous subsection, it was seen that the widths of correlations from these earlier times should be $\gtrsim 2$ units of relative rapidity. Any features observed with finer structure than these scales can probably be attributed to the continuous sources at times $\tau>\tau_0$. 

In this section we compare two correlations, one using the lattice equation of state, and one using an equation of state with constant speed of sound. In both cases, were present the correlations at a time when the temperature has fallen to $T=155$ MeV. Because experiments measure collisions with a range of centralities, the time at which the temperature reaches 155 MeV can vary. For the first case we consider $\tau_f=12$ fm/$c$, corresponding to a more central collision, and for the second case we consider $\tau_f=6$ fm/$c$, corresponding to a mid-central collision. For this grossly simplified picture, with no transverse flow, the evolution of the temperature can be taken from entropy conservation, $s(\tau)=s(\tau_f)\tau_f/\tau$. Given the entropy density, one then knows the temperature, and therefore the sources. For these cases we consider only the sources for $\tau>\tau_0=1$ fm/$c$.  

\begin{figure}
\centerline{\includegraphics[width=0.45\textwidth]{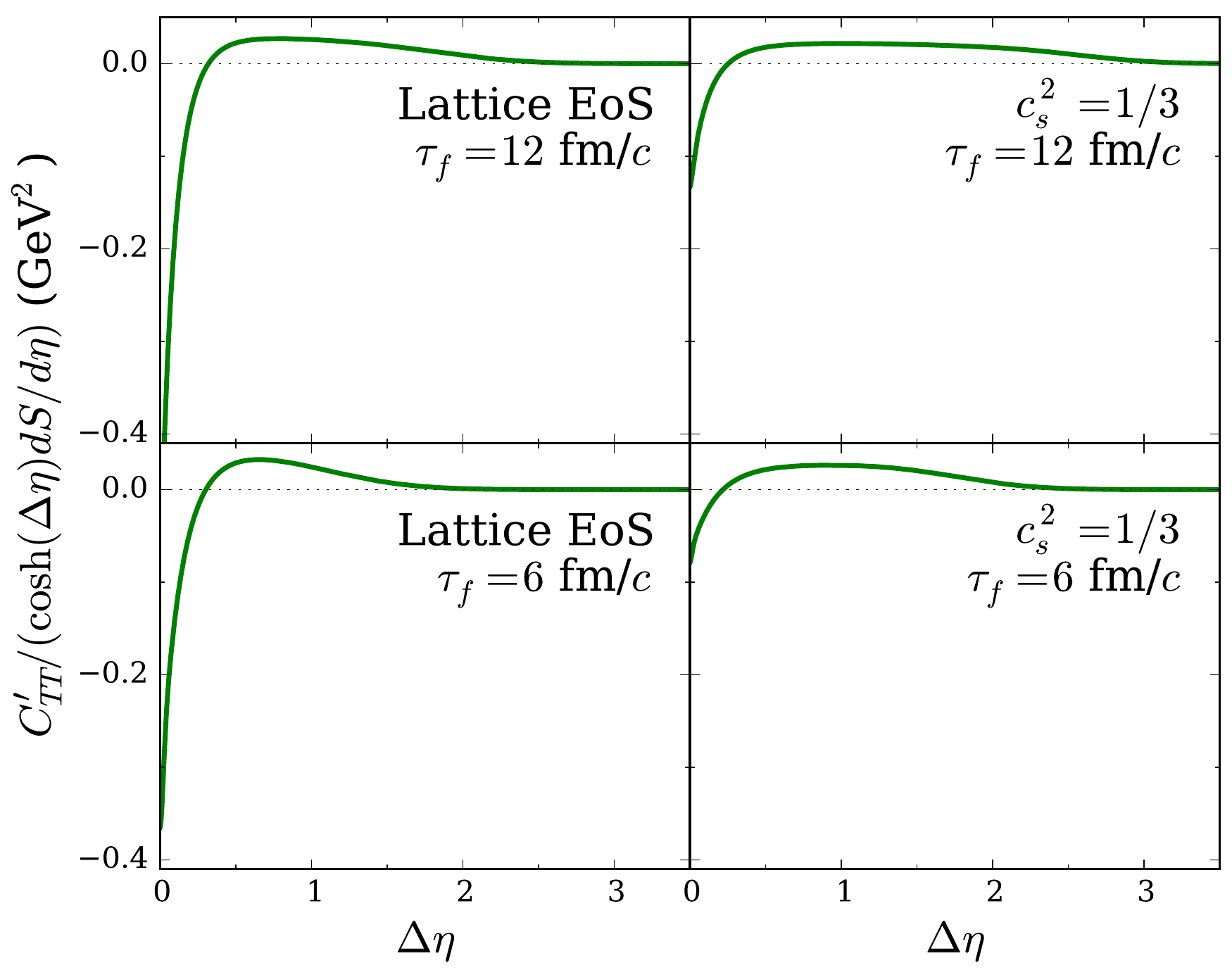}}
\caption{\label{fig:eteteos}(color online) 
Transverse energy correlations, scaled by the entropy per rapidity, as driven by the continuous source for $\tau>1$ fm/$c$, according to the self-consistent evolution equations for the correlations, Eq. (\ref{eq:masteretet}). Results are shown for both the lattice equations of state (left-side panels) and for a constant sound speed (right-side panels). The softer lattice equation leads to stronger negative correlations at small relative spatial rapidity. Correlations are displayed for a final temperature of 155 MeV, assuming that temperature was reached at 12 fm/$c$ (upper panels) to represent a more central collision, or at 6 fm/$c$ (lower panels) to represent a less central collision. For the more central case correlations extend to large relative rapidity.
}
\end{figure}

\begin{figure}
\centerline{\includegraphics[width=0.45\textwidth]{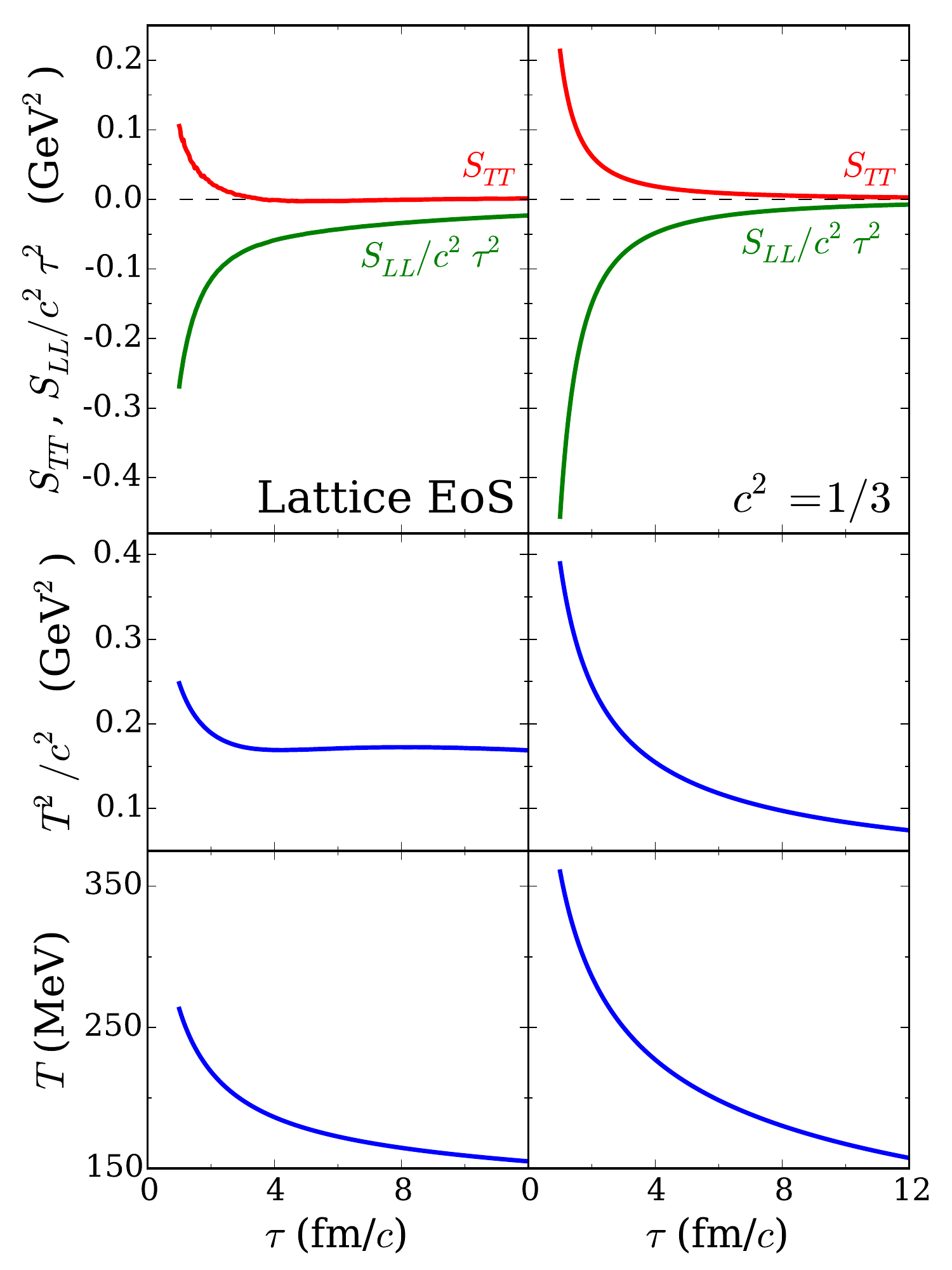}}
\caption{\label{fig:etetsource}(color online) 
Source functions for transverse energy correlations, $S_{TT}$, and for transverse momentum correlations, $S_{LL}$ are displayed as a function of time in the upper panels. The values are divided by the entropy per unit rapidity so that the values are independent of the transverse size. The transverse energy correlations, $C_{TT}$, are driven by the previous history of both $S_{TT}$ and $S_{LL}$ due to the equations of hydrodynamics which allow sound waves to be generate by either fluctuations in energy or momentum. The left-side panels show the results using the lattice equation of state, whereas the right-side panels show results for a constant speed of sound. The function $S_{LL}$ is scaled by $(c\tau)^{-2}$ due to the fact that it is defined with an extra factor of $\tau^2$, and is divided by $c^2$ because the strength of the energy and momentum in sound pulses varies by a factor $c$. The lattice equation of state has much stronger negative contributions at later times, which explains the stronger dips seen in the correlations of Fig. \ref{fig:eteteos}. The difference between $T^2$ and between $T^2/c^2$ for the two equations of state drive the source functions, as described in Table \ref{table:chiab}. }
\end{figure}

Figure \ref{fig:eteteos} shows the transverse energy correlator as calculated numerically from Eq.s (\ref{eq:masteretet}). Even though the models examined in this section are grossly oversimplified due to the lack of transverse expansion, one can take away several important conclusions. First, correlations of transverse energy binned by rapidity are driven by the evolution of the equilibrated local fluctuations of both the transverse energy and of the fluctuations of the longitudinal momentum. These correlations are linked because hydrodynamics links the equations of motion of perturbation of longitudinal momentum and of energy. Secondly, the spread of correlations due to a contribution from a source are very different from those for charge or for transverse momentum. In the other two cases, the spread is diffusive, whereas energy and longitudinal momentum are spread by waves. In the absence of viscosity the evolution involves slowly decaying delta functions, and once viscosity is included the spread is still far from Gaussian.

As was the case with charge correlations or with transverse momentum correlations, there are contributions from early times that are difficult, if not problematic, to understand. Energy and momentum correlations from early times are especially complicated by the role of jets and jet quenching. If however, one well understands the equation of state and viscosity, either from theory or from other measurements, these correlations might be used to constrain the contribution from early time, and perhaps to address critical questions about jet energy loss and equilibration in the first few tenths of a fm/$c$ of a collision.

\section{Projecting Correlations onto the Final State}
\label{sec:cascade}
Eventually, matter cools to a point where hydrodynamic equations of motion are no longer justified. Typically, this occurs as soon as the matter hadronizes. This is not so much due to long mean free paths, but do to the emergence of heavy hadronic states, mainly baryons, that have difficulty maintaining kinetic equilibrium with lighter species \cite{Pratt:1998gt,Sorge:1995pw}. At this point, one must consider the hundreds of hadronic species separately, and switch from hydrodynamic to microscopic models. Typically, these models involve simulations of individual hadrons scattering according to some prescription. The interface between the hydrodynamic models is typically a Cooper-Frye hypersurface, through which, on average, energy, momentum and charge are conserved. For a hadron species $h$, the number of particles with momentum $p$ that pass through a hyper surface element $d\Omega$ are 
\begin{eqnarray}
\label{eq:CooperFrye}
dN_h&=&\frac{d^3p}{(2\pi)^3E_p}(d\Omega\cdot p) f_h({\bf p})(2S_h+1),
\end{eqnarray}
where $f_\alpha({\bf p})$ is the phase space density of the given species, and is a function of the local temperature and flow velocity, but may also be corrected for distortions due to viscosity \cite{Pratt:2010jt,Molnar:2005gx,Dusling:2009df}. The conditions at the hypersurface are typically assumed to be those of a gas, and that emission of some species into a differential element $d^3p$ is uncorrelated with emission into any other cell once one knows $f$.

The challenge to address with correlations is how to adjust the phase space density to account for a perturbation in the flow of charge, momentum or energy that passes through the surface. First, we consider the correction to the charge/current density, $\delta j^\mu$. If one constrains the phase space density to not only maximize entropy for a fixed average energy, but also to fix the average four current $\delta j^\mu$, one would add four Lagrange multipliers to account for the four constraints, and in the frame of the fluid,
\begin{eqnarray}
f_h ({\bf p})&=&f_h^{\rm(0)}e^{\lambda_{a,\mu} \delta j^\mu_{h,a}({\bf p})},\\
\nonumber
j^\mu_{h,a}({\bf p})&=&q_{h,a}\frac{p^{\mu}}{E_p},\\
\nonumber
\delta f_h({\bf p})&\approx&f_h^{\rm(0)}({\bf p})~\lambda_{a,\mu}\delta j^\mu_{h,a}({\bf p}).
\end{eqnarray}
where $q_{h,a}$ is the charge of type $a$ on a hadron species $h$. The Lagrange multipliers $\lambda_{\mu,a}$ are chosen to reproduce the currents, and given that they are small in order to reproduce the small currents,
\begin{eqnarray}
\delta j^\mu_a&=&\sum_h \int\frac{d^3p}{(2\pi)^3} f_h^{\rm(0)}({\bf p}) q_{h,a}\frac{p^\mu}{E_p}\lambda_{\nu,b}q_{h,b}\frac{p^\nu}{E_p}\\
\nonumber
&=&-\chi^{\mu\nu}_{ab}\lambda_{\nu,b},\\
\nonumber
\chi^{\mu\nu}_{ab}&=&\sum_h\int \frac{d^3p}{(2\pi)^3} f_h^{\rm(0)}({\bf p}) q_{h,a}\frac{p^\mu}{E_p}\frac{p^\nu}{E_p}q_{h,b}.
\end{eqnarray}
Because we are considering a gas, where correlations are only between a particle and itself, $\chi$ can be associated with the thermal fluctuations of the charge and current,
\begin{equation}
\chi^{\mu\nu}_{ab}=\int d^3r~\langle j^{\mu}_a(0)j^{\nu}_b({\bf r})\rangle.
\end{equation}
One can now solve for $\lambda$,
\begin{eqnarray}
\lambda^\mu_a&=&(\chi^{-1})^{\mu\nu}_{ab}\delta j_{\nu,b},\\
\nonumber
(\chi^{-1})^{\mu\nu}_{ab}\chi_{\nu\gamma,bc}&=&g^\mu_{~\gamma}\delta_{ac}.
\end{eqnarray}
Finally, this gives $\delta f_h$,
\begin{equation}
\label{eq:deltaf}
\delta f_h({\bf p})= f_h^{\rm(0)}({\bf p}) q_{h,a}\frac{p_\mu}{E_p}(\chi^{-1})^{\mu\nu}_{ab}\delta j_{\nu,b}.
\end{equation}
One can test this solution and see that the emission of charge through the hyper-surface element $d\Omega$ is consistent,
\begin{equation}
\sum_h\int \frac{d^3p}{(2\pi)^3E_p}(d\Omega\cdot p)\delta f_h({\bf p}) q_a\frac{p^\mu}{E_p}=\delta j_a\cdot d\Omega.
\end{equation}

The matrix $\chi^{-1}$ is not as complicated as one might imagine. Given four Lorentz indices and perhaps three charges (up, down and strange or baryon number, isospin and strangeness), it appears as what might appear as a $12\times 12$ matrix. However, symmetry greatly reduces the complexity. The emission is linear in $d\Omega$, so one can consider the time-like, $d\Omega_0$, and space like, $d\vec{\Omega}$, contributions separately. From symmetry, in the matter frame there are no cross terms mixing the time-like and space-like Lorentz indices in $\chi$, and therefore not in $\chi^{-1}$. There can be cross terms between the spatial components when shear is present. For each set of indices $ij$ referring to two spatial indices in $\chi$, one could have separate $3\times 3$ matrices indexed by the charge indices. If one uses $\chi_{ab}$, to refer to the part of $\chi$ with Lorentz indices set to zero, and if for the spatial indices one considers $\chi_{ab}^{\rm(J)}$, Eq.s (\ref{eq:CooperFrye}) and (\ref{eq:deltaf}) can be rewritten so that 
\begin{eqnarray}
\delta dN_h&=&d\Omega_0 \frac{d^3p}{(2\pi)^3} f^{\rm(0)}({\bf p})q_{h,a}\chi^{-1}_{ab} \delta \rho_b,\\
\nonumber
&&+d\Omega_i \frac{d^3p}{(2\pi)^3E_p} f^{\rm(0)}({\bf p}) q_{h,a}\frac{p_i}{E_p}\left(\chi^{\rm(J)}\right)^{-1}_{ij,ab}
\delta j_{b,j}.
\end{eqnarray}
The first term, proportional to $d\Omega_0$ is the same as was used in \cite{Pratt:2012dz}. The second term requires calculating, then inverting, the matrix $\chi^{\rm(J)}$. If viscous effects are small, the matrix is diagonal and proportional to $\delta_{ij}$, so that
\begin{eqnarray}
\chi^{\rm(J)}_{ij,ab}&=&\delta_{ij}\chi^{\rm(J)}_{ab},\\
\nonumber
&=&\delta_{ij}\frac{1}{3}\sum_h\int\frac{d^3p}{(2\pi)^3}f_h^{\rm(eq)}q_{h,a}q_{h,b}\frac{|{\bf p}|^2}{E_p^2}.
\end{eqnarray}
The integral on the right-hand-side of the above equation is similar to the fluctuation used in the calculation of the conductivity in a Kubo relation, aside from a factor of the relaxation time divided by the temperature. Essentially, it is the current fluctuation of an equilibrated gas. If viscous effects are included, the integral still represents the current fluctuation of a gas, but not an equilibrated gas. Instead it is the current fluctuation in a gas with viscous distortions to the stress-energy tensor so that that matrix has off-diagonal elements.

Implementing this into a Monte Carlo can be straight-forward if one has already devised the means to generate the base distribution. For example in \cite{Pratt:2010jt}, one need only increase the probability of accepting a particle of a given species and momentum by a factor $(1+q_{h,a}\frac{p_\mu}{E_p}(\chi^{-1})^{\mu\nu}_{ab}\delta j_{\nu,b})$ as seen in Eq. (\ref{eq:deltaf}). This could be done by a keep-or-reject method, and can be helped by knowing that the maximum magnitude of the factor is bounded by knowing that ${\bf p}/E$ is always less than unity.

\section{Strategies and algorithms for evolving and projecting correlations}
\label{sec:algorithms}

There are numerous ways to implement the approaches presented here, or to implement approaches based on thermal noise \cite{Young:2014pka,Kapusta:2014dja,Kapusta:2012sd,Ling:2013ksb}. First, it is important to stress that the treatments based on thermal noise should give identical results as to those presented here if the mesh size used in the thermal noise calculations are small. This equivalency is demonstrated in the appendix. In thermal noise calculations, only single-particle distributions are evolved. The noise has the effect of generating both the peak of the correlation at $r=0$, and the correlation at $r\ne 0$. In contrast, the methods presented here assume there is a peak at $r=0$, and that its strength is consistent with local equilibrium. The calculations of the appendix show that the noise approach will also lead to peaks that are created nearly instantaneously, with the characteristic time scale being $a^2/D$, where $a$ is the mesh size and $D$ is the diffusion constant. 

An important issue in choosing an approach involves considering how one would treat the interface of the hydrodynamic codes with a microscopic simulation. Projecting the correlations can be rather tricky. If one ignores fluctuations and correlations and creates particles independently at the hyper-surface, as would be reasonable for an equilibrated gas, one still has fluctuations and correlations associated with the finite particle number. In fact these correlations reproduce those of an uncorrelated gas, which, according to lattice calculations \cite{Borsanyi:2014ewa}, is justified for a static equilibrated system in the temperature range, $\approx 155$ MeV, where interfaces are typically implemented. These correlations, are basically those of a particle with itself. In noise-based calculations these correlations appear as the very local, delta-function-like correlations on the scale of the hydrodynamic mesh. One should be careful to avoid double-counting of these fluctuations. If the correlations are fully confined to the hyper-surface element, one could imagine simply never creating more than one particle per element when creating particles to avoid projecting such correlations onto the simulation, and thus avoid double counting. In practice, this may be difficult because the hyper-surface mesh does not necessarily line up with the hydrodynamic mesh due to the fact that the hyper-surface moves relative to the hydrodynamic mesh, or to the fluid. Thus the advantage of noise-based calculations is that one can simply generate events from the noisy hydrodynamic calculation, then perform correlations the same as experiments would. The disadvantage is in avoiding the double-counting mentioned above.

For the methods described in the previous sections of this paper, there is no problem with the double counting because only $C'$, which ignores the delta function contribution is considered. However, then one needs to put more thought in how to best generate $C'$ and how to project that onto a simulation. It should be emphasized that the approach used in Eq.s (\ref{eq:masteretet}) in Sec. \ref{sec:et} is untenable for a realistic three-dimensional simulation. In this approach, the evolution of the correlation functions was explicitly followed. This was reasonable for the case where there was no transverse expansion, and where boost-invariance was assumed. In this very simplified case the correlation was purely a function of the relative coordinate $\Delta\eta$, whereas once energy, momentum and charge move transversely, one must treat the problem as a function of both coordinates $r_1$ and $r_2$. The tenable way forward is to generate pairs of charge (using the term charge generally to refer to charge, energy or momentum) at each point in space time, consistent with the source functions. One would then propagate the two charges independently, and combine the two single-particle distributions, weighted by the source functions, to generate the correlations.

More precisely, one can consider three ensembles, or simultaneous evolutions of the same initial conditions. The ensembles will be labeled I, II and III. Next, consider two charges $a$ and $b$ whose correlation one wishes to calculate. At each point in space-time, one would calculate the source for new correlation,
\begin{equation}
\delta s_{ab}\equiv S_{ab}d^4x.
\end{equation}
Next, one would add a charge $q_a$ to ensemble I with a random sign, and a charge $q_b$ to ensemble II.
\begin{eqnarray}
q_a&=&\pm \sqrt{\delta s_{ab}},\\
\nonumber
q_b&=&\delta s_{ab}/q_a.
\end{eqnarray}
The ensembles are then evolved with their new sources. The random sign, $\pm$, is different for each $d^4x$, with the volume perhaps being determined by the hydrodynamic mesh. In this way, the charges added to either ensemble are random, so the one-point distributions are not affected on average, but if one creates a correlation between ensemble I and ensemble II, one will reproduce the correlation $C'$. For cases where $a$ or $b$ refer to charges that don't average to zero, one must subtract an uncorrelated distribution to generate the correlation. For example, this is the case where $a$ or $b$ refer to baryon charge in lower energy collisions, or if $a$ or $b$ refer to energy. Ensemble III may be used for this purpose, or one might simply correlate an ensemble I calculation with calculations from completely different runs.. The correlation $C'$ is found be correlating ensemble I with ensemble II, whereas the correlation of a particle with itself, i.e. the equilibrium correlation, is generated by correlating ensemble III with itself. 

The method outlined above is noisy. There are a very large number of sources, one for each $d^4x$, but if each ensemble contains many fluctuations atop one another, each having been generated from a different point in space-time, there is noise from a contribution from a given $s_{ab}$ in ensemble I with a source in ensemble II from a different source point. Due to random phases, these should average to zero with sufficient statistics, but with finite statistics are noisy. 

In order to better understand the statistics necessary to overcome the noisy procedure outlined above, we present a simplified picture of $N$ sources, one for each point on some space-time mesh, and each with the same strength and width. One can imagine that each source creates balancing charges that spread in time, so that at the end of the hydrodynamic stage have spread according to a Gaussian distribution characterized by a Gaussian distribution. The two-particle correlation from each source point will have a form, $\sim s^i_{ab}\exp\{-\Delta y^2/4\sigma^2\}$. Here, the superscript $i$ refers to an individual source. One would then sum over the sources to get the net correlation. For our purposes we assume that the correlation sums to unit strength
\begin{equation}
\label{eq:simplegauss}
C_{ab}(\Delta\eta)=\sum_i s^i_{ab}e^{-\Delta y^2/4\sigma^2}=e^{-\Delta y^2/4\sigma^2}.
\end{equation}
If there are $N$ sources, there are $N$ contributions of strength $1/N$, but also $N^2/2$ contributions from two uncorrelated sources. The standard deviation of $N^2$ random numbers is of order $N$ multiplied which will then be of the same order as the actual signal. If one increases the number of sources by further dividing the source points, the contribution from each source scales as $1/N$, while the number of sources increases as $N$, so the signal should be independent of $N$ once $N$ is large. Further, the strength of the noise term should also remain unchanged. More precisely, one can consider a large range in rapidity $L_\eta$, where there are a large number of sources, $N$. Each source $i$ creates a single-particle distribution, $\sim e^{-(\eta-a_i)^2/2\sigma^2}$. If one assumes that the distribution of source points, $a_i$, is uniform within the range $L_\eta$, it is straight-forward to calculate the variance. For the case above where the correlation goes to unity at $\Delta\eta=0$, the standard deviation at any point is 
\begin{equation}
\label{eq:noise}
\langle \delta C(\Delta\eta)^2\rangle=\sqrt{(2\pi)^{1/2}\sigma/L_\eta}.
\end{equation}

\begin{figure}
\centerline{\includegraphics[width=0.4\textwidth]{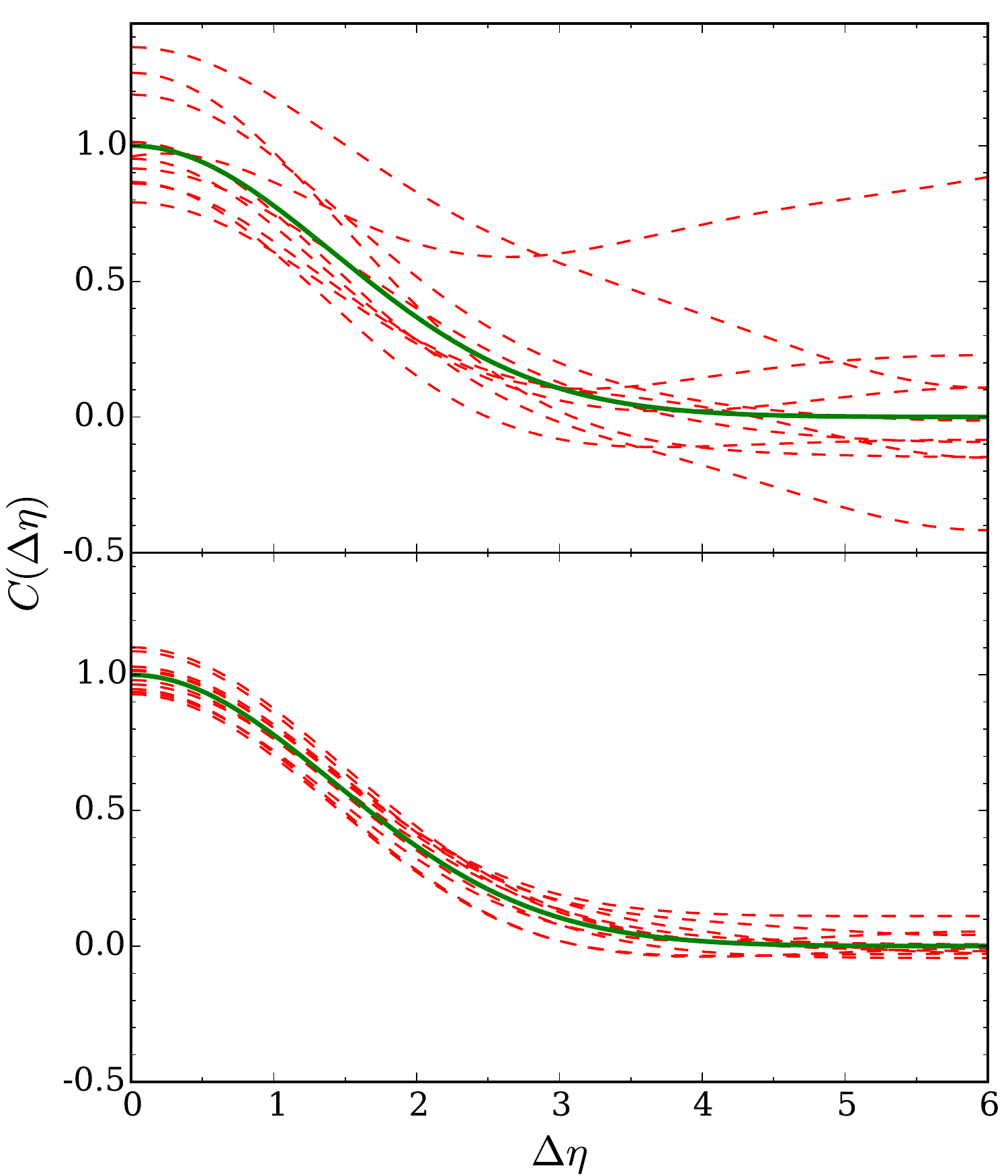}}
\caption{\label{fig:noise}(color online)
The solid green line shows Gaussian correlations coming from charge from source functions that have moved diffusively with a Gaussian spread of $\sigma=1$. Assuming a large number of sources, each contributing a with the same strength and width, we adjust the strength of the $N$ sources so that the resulting correlation in relative rapidity intersects the $\Delta\eta$ axis at unity. Including the noise by mixing contributions between different sources adds noise to the correlation. One such calculation gives noise as described in Eq. (\ref{eq:noise}). Averaging 10 such calculations is represented by the dashed red lines in the upper panels. The noise is highly correlated, so the averaging of 10 calculations is repeated several times to illustrate the variation of the curves due to a finite number of runs. In the lower panel 100 calculations are averaged, and the curves begin to reliably approach the correct answer. To obtain accuracy of better than 2 percent, roughly 1000 calculations, each with its own hydrodynamic evolution, are required.
}
\end{figure}

As stated in the previous paragraph, the uncertainty is independent of the number of sources used in the calculation, so further subdividing the sources is not helpful. However, increasing the rapidity range is useful because the source separated by several $\sigma$ do not both contribute to the same $\Delta\eta$. This is illustrated in \ref{fig:noise}. In this case $N=10^4$ sources of $\sigma=1$ were positioned randomly in a range $0<\eta<L_\eta=12$. The resulting distribution was convoluted with itself, with periodic boundary conditions applied to minimize edge effects. The two-particle distribution had contributions of each source with itself, the signal, and also contributions from different sources. On average, this second class of contribution should vanish. The process was repeated $n$ times and the $n$ correlations were averaged. For sufficiently large $n$, the correlations should approach the simple Gaussian of Eq. (\ref{eq:simplegauss}. However, the noise as described above makes each point uncertain by an amount $\sqrt{(2\pi)^{1/2}\sigma/nL_\eta}$, the noise mentioned above but divided by $\sqrt{n}$. The noise is highly correlated, at the same scale $\sigma$ as the signal. Calculations were performed for $n=10$ and for $n=100$. This was repeated ten times so that one could observe the spread of the result from finite statistics. For $n=1$ the uncertainty is approximately 0.46. The ten curves for $n=10$ vary by $0.46/\sqrt{10}$, but given the correlation between points, it is quite difficult to interpret the correlation. The procedure was repeated with $n=100$. The resulting ten curves are all much more representative of the true answer. If one repeats this procedure with $n=1000$, the uncertainty is a little more than one percent. Thus, obtaining calculations which fluctuate at the order of one or two percent require on the order of one thousand independent hydrodynamic runs, or more if the rapidity range $L/\sigma$ is less than the value 12 used here. Given the two ensembles for each set of sources, this would amount to 2000 runs.

One could avoid the noise mentioned above by performing the calculation for each source point independently. One could then choose the source points by Monte Carlo. Depending on how much the contribution from each source point varies in width and shape, this choice might require more or less than the thousand runs mentioned above. If each source point contributed with very similar shape, one would only need a few hydrodynamic runs, but if the source shape varied significantly, especially if some source contributed with opposite sign, one could need as many or perhaps more. Therefore, it is not immediately clear which strategy would be most efficient: hydrodynamic runs with all the source points contributing together, or isolating individual sources.

The second source of statistical noise in generating correlations comes from the generation of particles in the microscopic cascade. This mainly comes from the noise inherent to a smooth distribution. The strength of the noise is typically the number of uncorrelated particles, relative to a given particle, in a given bin compared to the number of correlated particles. Typically, for every particle, the net number of correlated particles from conservation laws is of order unity. For charge, this is clearly true, but it is also true for energy or momentum. If a particle is observed with a transverse momentum $p_x=1$ GeV/$c$, one knows exactly $-1$ GeV/$c$ can be found in the neighborhood, so the extra strength of the correlation tends to be similar to the strength of one particle. Thus, the statistical noise in a bin from correlations with other particles can be estimated by considering $N$ particles within a rapidity range whose size is determined by the spread of the correlation. The number of correlated pairs is then $N/2$, if the strength of the correlation is unity, and the number of uncorrelated pairs is $N^2/2$. Whereas the strength of the desired correlation is $\approx N/2$, and has minimal noise, the noise coming from uncorrelated pairs is then $\sqrt{N^2/2}$. This suggests that if one uses one event (or set of ensembles) the statistical noise will be roughly the size of the signal. If one wishes to reduce the relative size of the noise to the signal to one percent, one needs roughly $10^4$ events. Thus, when comparing low multiplicity events to high-multiplicity events, the statistical needs for a given accuracy tend to be independent of the multiplicity, or centrality. This is in contrast to considerations of noise for one-particle observables. In that case you need fewer events for high multiplicity to obtain the same signal to noise ratio. As stated above, the noise associated with finite particle number can be brought down to acceptable levels by evaluating thousands or perhaps tens of thousands of events. However, it should be emphasized that this does not require rerunning the hydrodynamics for each event. The microscopic simulations can be much faster than the hydrodynamic solutions, perhaps by factors of thousands or more. Thus, one could solve three hydrodynamic simulations, corresponding to the three ensembles above. Then one could repeat the process $n$ times. Calculating $C'$ could involve mixing any of the calculations for ensemble $I$ with those from ensemble $II$, thus leading to $n^2$ convolutions from the two ensembles. If one needed $10^4$ convolutions to reduce the noise, a choice of $n=100$ would suffice. 

The remaining correlation not included in $C'$, that of a particle with itself, must be calculated separately. For this correlation, one need only consider correlation within a smooth distribution. This also suffers from the same noise as described above. However, one need only perform one smooth hydrodynamic simulation, as opposed to the hundreds or thousands mentioned above. From that one hydrodynamic event, one would then perform many microscopic simulations to reduce the statistical noise.

If collisions within this simulation are ignored, and only decays are considered, one can eliminate nearly all the statistical noise by only considering the correlation of a particle with itself, or the correlation between the decay products of the same primary particle. This can be easily tracked if there are no collisions. If collisions are included, energy, momentum and charge are transferred approximately a  half dozen times for particles as they exit the collision region. This is sufficiently complicated that it is probably best to simply bin the correlations between all tracks, which brings back noise at the same level as discussed above, which would then require thousands, or perhaps even tens of thousands, of samplings. However, because smooth distributions are being used, one could avoid rerunning the hydrodynamics codes that many times, and given that the CPU demands of a microscopic simulation are far smaller than those of a hydrodynamic simulation, this could be rather tractable. 

Finally, if the particular analysis is entirely diffusive in nature, one might model the evolution of charge as a random walk atop the hydrodynamic background, similar in spirit as to what was done in Sec. \ref{sec:px}. By adjusting the collision time, and randomizing the single charge's movement at each ``collision'', one can mimic the diffusion equation. And by fixing the velocities so that no charges move faster than the speed of light, one can maintain causality. For considerations of charge moving through the quark-gluon plasma, this is probably the most reasonable approach. This strategy was also used for transverse momentum correlations in Sec. \ref{sec:px}, but once transverse expansion is included the evolution of transverse momentum mixes in energy and longitudinal momentum. The equations become coupled and non-diffusive in nature.

The strategies proposed here are mainly meant as suggestions, with the goal of illustrating the issues involved in generating correlations at a desired level of accuracy. As discussed above, one can apply a variety of tactics that can be formed into a strategy to optimize the analysis. Even though the best strategy is not known, it is clear that one can create excellent statistics without having to match the experiments by running as many hydrodynamic events as the experiment measures, which could be in the hundreds of millions or more. 

\section{Summary}
\label{sec:summary}
Correlation measurements from heavy-ion collisions have the potential to illuminate numerous fundamental properties of heavy-ion collisions. Energy fluctuations are related to the specific heat, momentum fluctuation to the temperature, and charge fluctuations to the charge susceptibility matrix. The challenge for phenomenologists is to understand how properties calculated in idealized systems based on the grand canonical ensemble, e.g. lattice gauge theory, can be linked to final-state properties of a finite-lived system where charge, energy and momentum are manifestly conserved.

The field has seen significant theoretical progress in the past few years. In \cite{Pratt:2012dz} and \cite{Ling:2013ksb} it was shown how the changing charge susceptibility served as sources for correlations of conserved charges. By observing the final-state separation of those charges, one could gain insight into the chemical history of the evolving collision \cite{Pratt:2015jsa}. Causal treatments have been introduced that better satisfy relativistic constraints \cite{Gavin:2016jfw,Kapusta:2014dja}. Correlated thermal noise in hydrodynamic treatments has been shown to provide a means for calculating charge correlations \cite{Ling:2013ksb,Kapusta:2014dja} by inserting noisy currents into the numerical evolution. Similar techniques have just begun to be applied to the calculation of correlations in energy and momentum, by adding noise into the spatial components of the stress-energy tensor \cite{Kapusta:2012sd,Young:2014pka}. 

In this paper we have provided a consistent perspective from which to consider and calculate correlations at the two-point level and connect them to fundamental properties for idealized systems such as susceptibilities or the specific heat. First, we show how the source for correlations in a dynamic heavy-ion environment is related to the rate-of-change of the susceptibility or fluctuation, where the fluctuations are calculated in the idealized environment of the canonical ensemble.For charge correlations, the source was given by the rate of change of the charge susceptibility or charge fluctuation in an idealized system. For transverse momentum, it was the rate of change of the temperature, and the source for transverse-energy correlations was the rate of change of the specific heat.  At the source, the correlations are perfectly local in coordinate space, but then spread with time. Here, we have compared and contrasted the way in which correlations then spread depending on the specific quantity: charge, energy or momentum. By considering the grossly over simplified picture of a Bjorken expansion with no transverse flow, we were able to perform calculations, sometimes analytically, to illustrate what properties of the matter determine the dynamical spreading of the correlations. For charge correlations, this was the diffusion constant for the specific charge. For transverse energy, it was the viscosity, and for transverse energy it was a combination of the speed of sound and viscosity.  In Sec. \ref{sec:algorithms}, strategies were presented to avoid double counting the auto-correlations of the hadron gas at the hydrodynamic interface with either the vacuum or with microscopic descriptions. Several issues were discussed here, but the decision on what may be the most efficient means for calculating correlations could not be determined without more study.

The paradigm presented here at first seemed simple. One identifies the conserved charges, then produce balancing charge pairs according to the rate of change of the local charge fluctuation or susceptibility. For charges, this seems rather straight-forward. The pairs are produced as point sources, which then simply diffuse separately. When the ``charges'' were energy or momentum, the situation was more complex, because energy and momentum evolve in tandem in a hydrodynamic treatment. Here, that was illustrated in a simple Bjorken (boost-invariant) treatment that neglected transverse expansion. In that case, the evolutions of longitudinal momentum and of energy had to be performed simultaneously. Meanwhile, transverse momentum correlations could be considered separately. In a realistic calculation, the inclusion of transverse dynamics would couple the entire stress-energy tensor, and one could not treat the transverse momentum correlations separately. Further, if one were to consider a system at finite charge density, such as a lower-energy heavy-ion collision, the transports of energy and of charge could no longer be considered independently. Thus, although the paradigm can be extended from the simple considerations of charge correlations in \cite{Pratt:2012dz}, it is clear that this is a substantial undertaking where the correlations of multiple conserved quantities would need to be considered simultaneously. In addition to encoding more sophisticated equations for the correlated transport, one may need to analyze thousands of hydrodynamic runs, differing only in the random sources. The development of efficient algorithms, for not only evolving the correlations but for projecting them through the interface between hydrodynamics and a microscopic hadronic cascade, is crucial

Even if the challenges enumerated in the previous paragraph and throughout this paper are met, the method remains limited in scope. The methods presented here, as well as those based on locally correlated noise, cannot account for equilibrated correlations of finite size. If one were to simulate a static system, the methods here would tend toward perfectly local, delta-function-like, correlations. The strengths would indeed be consistent with the susceptibility, but the equations presented here could not describe the size or growth of correlations that grow over large distances such as those near a critical point. It was shown in the appendix that noise-based pictures ultimately produce correlations with a length scale defined by the hydrodynamic mesh, and that the time that those short-range correlations adjust to the equilibrium levels is also given by the mesh size. If a system goes through phase separation, or even is close to a critical point, correlations build over large distances and the entire approach needs to be reconsidered.

The simple systems studied here were sufficient to illustrate several crucial points. Changes in the susceptibilities or in the specific heat indeed manifest themselves as sources and have clear measurable consequences. If the equation of state softens during the expansion, the speed of sound plummets and the specific heat peaks. This provides a negative source for energy correlations, whereas a drop in the specific heat would then produce a similar positive source for such correlations. Indeed, the results of Sec. \ref{sec:et} demonstrated a sensitivity to the equation of state. However, the effect was difficult to interpret because of the mixture of effects from the source for longitudinal momentum. The source for longitudinal momentum was shown to be related to changes in the product of the temperature and time. This factor of the time was due to the longitudinal expansion. Once one simultaneously considered both longitudinal momentum, the resulting behavior of correlations in transverse energy was complicated. Not only were there two sources of different character, but the hydrodynamic evolution mixed the responses. The correlations in transverse linear momentum, $\langle\delta P_x(0)\delta P_x(\eta)\rangle$ studied in Sec. \ref{sec:px}, were found to be significantly simpler. The source for the correlation was found to be related to the temperature, but without the factors of time that appeared for the longitudinal momentum. The evolution was found to be diffusive, with the diffusion constant being given by the viscosity. For this simple case, these correlations did not mix with those for energy or longitudinal momentum. The results of Sec. \ref{sec:px} show that one can indeed constrain some combination of the equation of state and viscosity. The final correlations in transverse momentum  differed substantially for different equations of state and for different viscosities.

All these correlations suffer from the same plague, that the initial correlation, in place when the system first thermalizes, is unknown. For energy and momentum correlations this is especially difficult because the role of jets and jet quenching is difficult. The net strength of this part of the correlation is determined by conservation laws, but the initial spread of such correlation is unknown. The subsequent evolution will smear out this contribution over at least a unit of rapidity, so the effect of sources from later times can be separated to some degree. However, the spread is not so large that the effects of initial correlation can be neglected. If one were to assume a form for the equation of state and viscosity, these measurements could perhaps shed light on the initial dynamics of energy and charge deposition. 

The varied ideas, calculations and strategies presented here by no means represent a conclusion. Implementing these ideas, or similar visions, requires significant development of algorithms and codes for hydrodynamics, for microscopic hadronic simulations, and for the interface. Hopefully, this study and discussion will help serve as a launching pad for moving forward. Progress will be difficult, but tenable, and the rewards are high. 

\appendix*

\section{The projection paradox}
As stated above, in the limit of large $\tau_0$ expansion is negligible and the response to a point-like perturbation is two receding plane waves. This may seem surprising, given that the response to a point-like perturbation is also a spherical wave. Reconciling the paradox involves understanding how a spherical plane wave, which clearly has energy in the region $|z|<ct$, takes the form $\delta(z-ct)\pm\delta(z+ct)$ when integrated over the $x-y$ plane. The solution to the paradox comes from looking in detail at the spherical solution. If $\epsilon$ is the energy density and $\pi_r$ is the radial momentum density, the equations of motion for a delta function like source at $r=t=0$ in a non-expanding medium are
\begin{eqnarray}
\partial_t\epsilon=\partial_r\pi_r-+2\pi_r/r&=&E\delta^3(r)\delta(t),\\
\nonumber
\partial_t\pi_r-c^2\partial_r\epsilon&=&0.
\end{eqnarray}
The solution for these boundary conditions are:
\begin{eqnarray}
\epsilon&=&-\frac{E}{4\pi r}\delta'(r-ct),\\
\nonumber
\pi_r&=&-\frac{c^2Et}{4\pi r^2}\delta'(r-ct),
\end{eqnarray}
and can be found in textbooks \cite{LandauLifshitzFluidMechanics}. Here, $\delta'$ is the derivative of the delta function. 

The $\delta'$ terms can be thought of as two delta functions, one with positive energy density, and one with negative energy density, separated by some small radial distance $\Delta$, 
\begin{eqnarray}
-\delta'(r-ct)&=&\frac{1}{\Delta}\left\{\delta(r-ct-\Delta)-\delta(r-ct)\right\},\\
\nonumber
4\pi\int r^2dr~\epsilon&=&\frac{E}{\Delta}\left\{(ct+\Delta)-ct\right\}=E.
\end{eqnarray}
The total energies of the outer/inner shells, at $r+\Delta$ and $r$, are $(r+\Delta)E/\Delta$ and $-rE/\Delta$. The energy of each individual shell is linear in $r$ and increases linearly with time because $r=ct$, but the net energy of the two shells is fixed at $E$. The energy in a solid angle corresponding to a projection for a given $dz$ is
\begin{eqnarray}
dE&=&\frac{2\pi E}{4\pi}\left\{(ct+\Delta)d\cos\theta_+-(ct)d\cos\theta_-\right\},
\end{eqnarray}
where the solid angles subtended by the two shells are $2\pi d\cos\theta_+=2\pi dz/(ct+\Delta)$ and $2\pi d\cos\theta_-=2\pi dz/ct$ respectively. Thus,
\begin{eqnarray}
\frac{dE}{dz}=0,~{z\ne r}.
\end{eqnarray}
The projected energy density thus cancels from the two spheres because the smaller sphere, even though it has slightly less energy, subtends slightly more solid angle in a cut defined by $dz$. This can also be seen more formally in cylindrical coordinates, $\rho^2+z^2=r^2$,
\begin{eqnarray}
\frac{dE}{dz}&=&-\frac{E}{4\pi}\int 2\pi\rho d\rho~ \frac{1}{r}\delta'(r-ct),\\
\nonumber
&=&-\frac{E}{2}\int \rho d\rho~ \frac{\partial\rho}{\partial r}\frac{1}{r}\frac{d}{d\rho}\delta(r-ct),\\
\nonumber
&=&-\frac{E}{2}\int_0^\infty d\rho~ \frac{d}{d\rho}\delta(\sqrt{\rho^2+z^2}-ct)\\
\nonumber
&=&\frac{E}{2}\left\{\delta(z-ct)+\delta(z+ct)\right\}.
\end{eqnarray}
Thus, the point source projects to two receding plane waves when integrating over $x$ and $y$.

\section{Equivalence to Derivations Based on Thermal Noise}
\label{sec:thermalnoise}

Equivalent derivations of the main expressions for correlations presented here can be derived by considering single-particle currents in the presence of thermal noise \cite{Kapusta:2012sd,Ling:2013ksb}. For completeness, we include variations of those derivations along with a discussion of why the perspectives lead to the same expressions.

In calculations incorporating thermal noise, the charge densities and currents are split into two pieces, a diffusive piece and a noisy piece that averages to zero,
\begin{eqnarray}
\label{eq:jsn}
\vec{j}(x)&=&\vec{j}^{\rm(d)}+\vec{j}^{\rm(n)}.\\
\nonumber
\vec{j}^{\rm(d)}(x)&=&-D\nabla\delta\rho(x),\\
\end{eqnarray}

The noisy current, or that part that differs from the diffusive contribution, averages to zero, but is correlated at short distances and relative times,
\begin{equation}
\label{eq:kubojj}
\langle j_i^{\rm(n)}(x_1)
j_k^{\rm(n)}(x_2)\rangle=2\sigma T\delta_{ik}\delta^4(x_1-x_2),
\end{equation}
where $\sigma$ is the conductivity. The assumption of a delta function is reasonable if the contributions of the correlations in the Kubo relation,
\begin{equation}
\sigma=\frac{1}{6T}\int_{-\infty}^\infty dx_0\int d^3x~ \langle\vec{j}(x)\cdot\vec{j}(0)\rangle,
\end{equation}
are confined to small relative distances and times.

To consider the rate of change of the correlation function, we take the difference of the correlation between two times separated by $\Delta t$,
\begin{eqnarray}
\nonumber
C(\vec{x}_1,\vec{x}_2,t)&\equiv&\langle\delta\rho(\vec{x}_1,t)\delta\rho(\vec{x}_2,t)\rangle,\\
\nonumber
\delta\rho(\vec{x}_1,\vec{x}_2,t+\Delta t)&=&
\delta\rho(\vec{x}_1,\vec{x}_2,t)-\int_{t}^{t+\Delta t}dt'~\nabla\cdot(-D\nabla\delta\rho(\vec{x},t')+\vec{j}^{\rm(n)}(\vec{x},t'),\\
\nonumber
C(\vec{x}_1,\vec{x}_2,t+\Delta t)&=&C(\vec{x}_1,\vec{x}_2,t)+D\Delta t(\nabla_1^2+\nabla_2^2)C(\vec{x}_1,\vec{x}_2,t)\\
\nonumber
&-&\int_{t}^{t+\Delta t}dt'dt''~\langle(\nabla\cdot\vec{j}^{\rm(n)}(\vec{x}_1,t'))(\nabla\cdot\vec{j}^{\rm(n)}(\vec{x}_2,t''))\rangle\\
\nonumber
&=&C(\vec{x}_1,\vec{x}_2,t)+\Delta t\left\{D(\nabla_1^2+\nabla_2^2)C(\vec{x}_1,\vec{x}_2,t)-2\sigma T(\nabla_1\cdot\nabla_2) \delta^3(\vec{x}_1-\vec{x}_2)\right\},\\
\label{eq:dCdtgeneral}
\partial_t C(\vec{x}_1,\vec{x}_2,t)&=&D(\nabla_1^2+\nabla_2^2)C(\vec{x}_1,\vec{x}_2,t)-2\sigma T(\nabla_1\cdot\nabla_2) \delta^3(\vec{x}_1-\vec{x}_2).
\end{eqnarray}
The third line invoked the Kubo relation, Eq. (\ref{eq:kubojj}), and it has been assumed implicitly that it is sufficient to consider time steps $\Delta t$ larger than the characteristic short correlation time used to motivate the use of delta functions in the Kubo relations.

One can relate the conductivity $\sigma$ to the diffusion constant by equating the electric field energy, $x\rho(x)E_x$, to a gradient in the chemical potential, $-x\rho(x)\partial_x\mu$,
\begin{eqnarray}
\vec{j}&=&\sigma\vec{E}=-\sigma\partial_x\mu,\\
\nonumber
&=&-\sigma\frac{\partial_x\rho}{\partial_\mu \rho}.
\end{eqnarray}
Given that the susceptibility is $\chi=T\partial_\mu\rho$,
\begin{eqnarray}
\vec{j}&=&-\frac{T}{\chi}\partial_x\rho=-D\partial_x\rho,\\
\nonumber
D\chi&=&\sigma T.
\end{eqnarray}

Eq.s (\ref{eq:dCdtgeneral}) can be simplified when the correlation function is projected in terms of the relative coordinate, $\vec{r}=\vec{x}_1-\vec{x}_2$. Combined with the definition of the average coordinate $\vec{R}=(\vec{x}_1+\vec{x}_2)/2$,
\begin{eqnarray}
\nabla_1&=&\nabla_r+\nabla_R/2,~\nabla_2=-\nabla_r+\nabla_R/2,\\
\nonumber
\nabla_1^1+\nabla_2^2&=&\frac{1}{2}\nabla_R^2+2\nabla_r^2,\\
2\nabla_1\cdot\nabla_2&=&\frac{1}{2}\nabla_R^2-2\nabla_r^2.
\end{eqnarray}
Inserting these into Eq. (\ref{eq:dCdtgeneral}) one can solve for the correlation function in terms of the relative coordinate,
\begin{eqnarray}
\label{eq:Cofroft}
C(\vec{r},t)&=&\frac{1}{V}\int d^3R\langle \delta\rho(\vec{R}+\vec{r}/2)\delta\rho(\vec{R}-\vec{r}/2)\rangle,\\
\nonumber
\partial_t C(\vec{r},t)-2D\nabla_r^2 C(\vec{r},t)&=&-2D\chi\nabla^2\delta^3(\vec{r}).
\end{eqnarray}
At equilibrium, $\partial_tC=0$, and one finds the correlation at equilibrium,
\begin{eqnarray}
C(\vec{r},t)=\chi\delta^3(\vec{r}).
\end{eqnarray}

The expressions derived in this appendix thus far can be found in text books \cite{LandauLifshitzFluidMechanics}. The next step is to understand how Eq. (\ref{eq:Cofroft}) leads to an understanding of the dynamics of how equilibrium is attained. To that end we express the delta function as the limit of a thin Gaussian of width $a$. At equilibrium, the correlation function is
\begin{equation}
C^{\rm(eq)}(r)=\frac{\chi}{(2\pi a^2)^{3/2}}e^{-r^2/2a^2},~~~a\rightarrow 0.
\end{equation}
One can then find the evolution of the difference between the true correlation and the equilibrium correlation function as
\begin{eqnarray}
C'(r,t)&\equiv&C(r,t)-C^{\rm(eq)}(r),\\
\nonumber
\partial_t C'(r,t)-2D\nabla^2 C'(r)&=&0.
\end{eqnarray}
Here, we have assumed that $C^{\rm(eq)}$ is independent of time, i.e. $\chi$ is fixed. For the initial condition $C(r,t=0)=0$, the solutions to the equations of motion are
\begin{eqnarray}
C'(r,t=0)&=&-\frac{\chi}{(2\pi a^2)^{3/2}}e^{-r^2/2a^2},\\
\nonumber
C'(r,t)&=&-\frac{\chi}{(2\pi [a^2+2Dt])^{3/2}}e^{-r^2/2[a^2+2Dt]^2},\\
\nonumber
C(r,t)&=&\frac{\chi}{(2\pi a^2)^{3/2}}e^{-r^2/2a^2}-\frac{\chi}{(2\pi [a^2+2Dt])^{3/2}}e^{-r^2/2[a^2+2Dt]^2}.
\end{eqnarray}
The characteristic time for returning to equilibrium is $\tau=a^2/D$, and as $a\rightarrow 0$, the system is restored to equilibrium instantaneously near the origin because the infinite peak in $C'$ at $r=0$,
\begin{equation}
C'(r=0,t)\sim-\frac{1}{[a^2+2Dt]^{3/2}},
\end{equation}
becomes finite in an infinitesimal time as $a\rightarrow 0$. Once $t>a^2/2D$, the correlation $C'$, which balances the peak at $r=0$, is smooth and spreads out diffusively. In a numerical treatment of the problem the finiteness of $a$ is defined by the grid size for the hydrodynamic treatment, and for small grid sizes, this peak should rapidly approach the equilibrium value.

In the limit $a\rightarrow 0$, described above, the treatment of the correlation in terms of noise then exactly replicates the expressions applied throughout the body of this paper. This is because the peak of the short range correlation then exactly matches the equilibrium value determined by $\chi$ and the part of the correlation away from $r=0$ spreads diffusively. Once can consider the problem where in a time interval $\Delta t$ the system changes such that $\chi$ changes by $\Delta\chi$. The correlation $C'$, then changes by an amount,
\begin{eqnarray}
\Delta C'(r,t+\Delta t)&=&-\frac{\Delta\chi}{(2\pi a^2)^{3/2}}e^{-r^2/2a^2},
\end{eqnarray}
which then spreads diffusively. The source function for the diffusion is thus $\delta^3(\vec{r})d\chi/dt$ in the limit $a\rightarrow 0$.

Thus, in the limit that the noise is correlated perfectly locally in space-time, the approaches become identical and has many of the same drawbacks. Mainly, if the correlation should have a significantly large extent, or if the local dynamics are constrained physically by some other means such as the finite time required to create certain species, the approximations are no longer justified.

\begin{acknowledgments}
This work was supported by the Department of Energy Office of Science through grant number DE-FG02-03ER41259.
\end{acknowledgments}


\end{document}